\documentclass{article}

\usepackage{arxiv}
\usepackage{framed,multirow}

\usepackage[utf8]{inputenc} 
\usepackage[T1]{fontenc}    
\usepackage{hyperref}       
\usepackage{url}            
\usepackage{booktabs}       
\usepackage{amsfonts}       
\usepackage{nicefrac}       
\usepackage{microtype}      
\usepackage{lipsum}		
\usepackage{graphicx}
\usepackage{natbib}
\usepackage{doi}

\usepackage{amsmath}
\usepackage{amssymb}
\usepackage{latexsym}

\usepackage{wrapfig}
\usepackage{graphicx}
\usepackage{subfig}
\usepackage{dcolumn}

\begin{document}

\title{A quantitative comparison of physical accuracy and numerical stability of Lattice Boltzmann color gradient and pseudopotential multicomponent models for microfluidic applications}%

\author{ Karun P.N. Datadien \\
	Department of Applied Physics\\
	Eindhoven University of Technology\\
	5612 AZ Eindhoven \\
	\texttt{ k.p.n.datadien@tue.nl} \\
	\And
	Gianluca Di Staso \\
	Department of Applied Physics\\
	Eindhoven University of Technology\\
	5612 AZ Eindhoven \\
	\AND
	Herman M.A. Wijshoff \\
	Department of Mechanical Engineering\\
	Eindhoven University of Technology\\
	5612 AZ Eindhoven \\
	\And
           Federico Toschi \\ 
	Department of Applied Physics\\
	Eindhoven University of Technology\\
	5612 AZ Eindhoven \\
}

\date{\today}

\maketitle

\begin{abstract}

The performances of the Color-Gradient (CG) and of the Shan-Chen (SC) multicomponent Lattice Boltzmann models are quantitatively compared side-by-side on multiple physical flow problems where breakup, coalescence and contraction of fluid ligaments are important. The flow problems are relevant to microfluidic applications, jetting of microdroplets as seen in inkjet printing, as well as emulsion dynamics. A significantly wider range of parameters is shown to be accessible for CG in terms of density-ratio, viscosity-ratio and surface tension values. Numerical stability for a high density ratio $\mathcal{O}(1000)$ is required for simulating the drop formation process during inkjet printing which we show here to be achievable using the CG model but not using the SC model. In terms of physical accuracy, the CG model shows good agreement with analytical solutions for droplet oscillation and ligament contraction test-cases. The SC model is effective in accurately simulating ligament contraction, but less accurate in handling droplet oscillations. Rayleigh-Plateau instability simulations show that both the CG and SC models give physically realistic results when breakup occurs, although smaller satellite droplets quickly vanish in the SC model due to droplet mass evaporating into the ambient phase. Our results show that the CG model is a suitable choice for challenging simulations of droplet formation, due to a combination of both numerical stability and physical accuracy. We also present a novel approach to incorporate repulsion forces between interfaces for CG, with possible applications to the study of stabilized emulsions. Specifically, we show that the CG model can produce similar results to a known multirange potentials extension of the SC model for modelling a disjoining pressure, opening up its use for the study of dense stabilized emulsions.

\end{abstract}



\section{\label{sec:level1}Introduction\protect\\ }

The numerical modeling of multiphase/multicomponent fluids is still a challenge and the Lattice Boltzmann Method (LBM) has shown great potential in this field \cite{succi}. Several models for simulating multiphase/multicomponent flows using the LBM have been proposed over the last three decades, including the color gradient (CG) model \cite{gunstensen24}, the pseudopotential model \cite{SC}, the free-energy model \cite{swift31} and the mean-field model \cite{He13}. In the current work the CG  model - based on a two-species variant of the lattice gas automata model introduced in \cite{Rothman} - is compared to the classical pseudopotential model developed by Shan and Chen (SC) and first introduced in \cite{SC}. The aim of comparing the two models is to quantitatively characterize how well each model performs in the context of realistic flow problems involving challenging interface dynamics, where e.g. surface tension and disjoining pressure play a crucial role in the breakup of fluid ligaments and in the coalescence of droplets. Specifically, our main focus is to investigate the feasibility and limitations of each of the two models for accurately simulating the jetting of microdroplets \cite{WIJSHOFF2010}, a challenging industrial application where accurate modeling of multicomponent fluids is essential. During a typical jetting cycle, the ejected droplet is usually followed by a long attached tail/ligament. This ligament can either contract and coalesce with the main droplet or detach from the main droplet and contract into one or more detached satellite droplets. The droplet(s) formed during a jetting cycle will oscillate to some degree with a frequency that is analytically known \cite{Miller1968}. Besides LBM based models, there are several other popular approaches that could be considered to simulate such a jetting system, among which the front-tracking method \cite{fronttracking}, the volume-of-fluid (VOF) method \cite{VOF} and the level-set method \cite{levelset}. These methods are based on solving the macroscopic Navier-Stokes equations alongside with a technique to track the interface between different phases and apply interfacial tension \cite{Liu}. In the front-tracking method interface breakup does not automatically arise from numerical modeling of the interface, which necessitates manually rupturing the interface, according to an \textit{ad hoc} criterion, e.g. as described in \cite{TRYGGVASON2001708}, in order to model interface breakup physics. On the contrary, both VOF and level-set methods naturally capture breakup and coalescence of interfaces. The VOF method requires interface reconstruction for determining and applying the proper surface tension, which can be a computationally expensive operation and may not always be physically consistent \cite{YANG2006364}. It has also been shown in \cite{scardo} that both VOF and level set methods suffer from numerical instability around the interface when complex geometries are considered in combination with interfacial tension being the dominant force.

LBM based multiphase/multicomponent models have the advantage that coalescence and breakup events naturally arise from solving the mesoscopic level equations and no further manual intervention is required. Furthermore no additional computational resources are used for interface tracking, meaning that the computational load is independent of the amount of interfacial area present in the simulated system. A primary requirement of the model, in order to simulate ink-air systems such as the formation of microdroplets during inkjet printing, is numerical stability for a high density ratio between ink and air, typically in the order of thousands, $\mathcal{O}(1000)$. Ideally the numerical model should also support a wide range of viscosities and surface tensions that can be set independently from other parameters, since ink properties may vary depending on chemical composition. 

The cases presented in this paper aim at characterizing the performance of the CG and SC models for several physical phenomena relevant to inkjet printing. We investigate physical accuracy and/or stable parameter ranges for a series of classical benchmark cases plus some more complex multicomponent emulsion flow. Specifically we focus on the following six cases:
(1) A Laplace law test where we measure the accuracy in recovering the surface tension as expected according to the Young-Laplace law. We also quantify the spurious currents in the system, which may influence simulation results depending on their magnitude. (2) A droplet oscillation test where we measure the oscillation frequency of a droplet and compare it to the analytical solution reported in \cite{Miller1968}. (3) A viscous ligament contraction test where the measured contraction rate of a ligament is compared to the analytical approximation reported in \cite{Srivastava2013}. (4) A Rayleigh-Plateau (RP) instability test where we measure the size of two resulting droplets (main and satellite) after breakup of a cylindrical ligament. The results are compared to simulation results reported in \cite{slenderjet}.

Furthermore we introduce a novel method of applying a repulsion force at an interface for the CG model that can be, e.g., used to mimic the effect of surfactants. We demonstrate the applicability of our method for two use cases. Firstly the case (5) of droplets colliding with a repulsion force acting between the interfaces. We compare the behavior of the novel method presented in this paper - introduced in section \ref{sec:repforce} - to SC simulations with multirange potentials to implement repulsion (as presented in \cite{Benzi2009}). This is done by measuring the radius of the liquid bridge as a function of time during coalescence events. The other case (6) is that of emulsion mixing where we compare the SC double belt results with our CG repulsion force implementation by tracking the number of droplets in the system as a function of time.

\section{Multiphase/multicomponent methods}

In this section we introduce the numerical methods we employ throughout the paper. This includes the classical Shan-Chen model and the color gradient model with enhanced equilibrium distribution functions, as well as extensions to include a repulsion force for each method.

\subsection{The classical Shan-Chen model}\label{sec:SC}

In the classical Shan-Chen model, to simulate (multiphase or multicomponent) flows, a particle distribution function is evolved according to the lattice Boltzmann equation:

\begin{equation}
f_i(\vec{x}+\vec{c}\Delta t,t+\Delta t) = f_i(\vec{x},t) +  \Omega_i ( f_i(\vec{x},t))\Delta t
\label{eq:streamingSC}
\end{equation}

with $  f_i(\vec{x},t)  $ being the discrete particle distribution functions at lattice site $\vec{x}$ at time $t$. From now on we will use the standard convention that $\Delta t = 1$. 

The Bhatnagar-Gross-Krook (BGK) collision operator implements linear relaxation towards the equilibrium and has the form
\begin{equation}
\Omega_i ( f_i(\vec{x},t)) = - \frac{1}{\tau} [ f_i(\vec{x},t) -  f_i^{eq}(\rho(\vec{x},t),\vec{u}(\vec{x},t))] + S_i(\vec{x},t)
\label{eq:collisSC}
\end{equation}

with $\tau$ being a relaxation time and $S_i$ an optional source term used, e.g., to incorporate forcing. The equilibrium distribution functions, $f_i^{eq}$, are defined as

\begin{equation}
f_i^{eq}(\rho(\vec{x},t),\vec{u}(\vec{x},t)) =\rho w_i \left( 1 + \frac{3}{c^2} \vec{c}_i \cdot \vec{u} +\frac{9}{2 c^4} (\vec{c}_i \cdot \vec{u})^2 - \frac{3}{2 c^2}(\vec{u})^2  \right)
\label{eq:equiliSC}
\end{equation}

with density $\rho$ and (macroscopic) fluid velocity $\vec{u}$. The discrete velocities $\vec{c}_i$ used for the standard D3Q19 scheme are

\begin{equation}
   \vec{c}_i  = 
    \begin{cases}
      (0,0,0), & i = 1 \\
      (\pm1,0,0), (0,\pm1,0), (0,0,\pm1) & i = 2 - 7 \\
      (\pm1,\pm1,0), (\pm1,0,\pm1), (0,\pm1,\pm1) & i = 8 - 19
    \end{cases}
\label{eq:speeds}
\end{equation}

with the weights  set to the following values:

\begin{equation}
   w_i  = 
    \begin{cases}
      1/3, & i = 1 \\
      1/18 & i = 2 - 7 \\
      1/36 & i = 8 - 19
    \end{cases}
\label{eq:weights}.
\end{equation}

The density of the fluid is recovered by summing the particle distributions as 
\begin{equation}
\rho = \sum_i f_i
\label{eq:densitySC}.
\end{equation} 

The velocity $\vec{u} = (u_x, u_y, u_z)$ is evaluated as the first order moment of the distribution according to the relation

\begin{equation}
\rho \vec{u}  = \sum_i f_i \vec{c}_i 
\label{eq:totalMomentum}.
\end{equation}

For multi-phase simulations the interaction between two phases is included through the forcing term \cite{PhysRevE.49.2941} 

\begin{equation}
\vec{F} (\vec{x},t) = -G \psi(\vec{x},t) \sum_i w_i \psi(\vec{x}+\vec{c}_i,t)\vec{c}_i
\label{eq:interactionforceMP}
\end{equation}

where $G$ is the coupling parameter that determines the interaction strength, where $G > 0$ will result in repulsion forces and $G < 0$ results in attractive forces.  The pseudo-density function, $ \psi(\rho)$, is set as \cite{SC}

\begin{equation}
 \psi(\rho) = \rho_0 [1 - \exp(-\rho/\rho_0))]
\label{eq:pseudopotential}
\end{equation}

where $\rho_0$ refers to a reference density. The application of a force, including the interaction between two fluids, is achieved, e.g., through the forcing scheme proposed by Guo \cite{SRIVASTAVA_2013,Guo}. The equilibrium velocity is shifted as follows:

\begin{equation}
\vec{u}^{eq}(\vec{x},t) = \frac{1}{\rho(\vec{x},t)} \sum_i \vec{c}_i f_i(\vec{x},t) + \frac{\vec{F}(\vec{x},t)}{2 \rho(\vec{x},t)}
\label{eq:guoforce}.
\end{equation}

The source term $S_i$ takes the form

\begin{equation}
S_i = w_i \left(1 -  \frac{1}{2 \tau} \right) \left( \frac{(\vec{c}_i - \vec{u})\cdot \vec{F}}{c_s^2} + \frac{(\vec{c}_i \cdot \vec{u})(\vec{c}_i \cdot \vec{F})}{c_s^4} \right)
\label{eq:source}
\end{equation}

with 

\begin{equation}
\vec{u}(\vec{x},t^*) = \frac{1}{\rho(\vec{x},t)} \sum_i \vec{c}_i f_i(\vec{x},t) + \frac{\vec{F}(\vec{x},t)}{2 \rho(\vec{x},t)}
\label{eq:guoforce2},
\end{equation}

where $t^*$ denotes a time after the collision step, but before the streaming step. 
For the simulations in this work where repulsion forces are also present, we use a multi-component variation of the SC model with the extension of a repulsion model discussed in \cite{Benzi2009}. The gist of the method is that there is a purely attractive ($a$) pseudo-potential force acting on the first ``Brillouin zone" (``belt 1") and a repulsive ($r$) force acting on both ``belt 1" and the second ``Brillouin zone" (``belt 2"). The force acting between different species ($X$) is repulsive and short-ranged only. Therefore, the force $\vec{F}_s$ acting on species $s$ takes the form

\begin{equation}
\vec{F}_s(\vec{x},t) = \vec{F}^a_s(\vec{x},t) + \vec{F}^r_s(\vec{x},t)  + \vec{F}^X_s(\vec{x},t)
\label{eq:benziforces}
\end{equation}

with

\begin{equation}
\vec{F}^a_s(\vec{x},t) = -G^a_s \Psi_s(\vec{x},t) \sum_{i \in belt1} w_i \Psi_s(\vec{x}+\vec{c}_i,t) \vec{c}_i
\label{eq:doublebeltattractiveforce},
\end{equation}

\begin{equation}
\begin{split}
\vec{F}^r_s(\vec{x},t) = & -G^r_s \Psi_s(\vec{x},t) \sum_{i \in belt1} p_i \Psi_s(\vec{x}+\vec{c}_i,t) \vec{c}_i  - G^r_s \Psi_s(\vec{x},t) \sum_{i \in belt2} p_i \Psi_s(\vec{x}+\vec{c}_i,t) \vec{c}_i 
\end{split}
\label{eq:doublebeltrepulsionforce}
\end{equation}

and 

\begin{equation}
\vec{F}^X_s(\vec{x},t) = -\frac{1}{(\rho^{(s)}_0)^2}\rho_s(\vec{x},t)  \sum_{s' \neq s}  \sum_{i \in belt1} G_{ss'} w_i \rho_{s'}(\vec{x}+\vec{c}_i,t) \vec{c}_i 
\label{eq:doublebeltrepulsionforce2}, 
\end{equation}

where we have $G_{ss'} = G_{s's}$ with $s' \neq s$, the cross-coupling constant between species $s$ and $s'$. Here $p_i$ are the associated forcing weight which we set as:

\begin{equation}
   p_i  = 
    \begin{cases}
      0 & i = 1 \\
      4/135 & i = 2 - 7 \\
      1/63 & i = 8 - 19 \\
      2/315 & i = 20 - 27 \\
      5/1512 & i = 28 - 33 \\
      1/945 & i = 34 - 57 \\
      1/1890 & i = 58 - 81 \\
      1/15120 & i = 82 - 93 
    \end{cases}
\label{eq:pweights}.
\end{equation}

\subsection{The color gradient model}
In the implementation of the CG model proposed here, there is one set of discrete distribution functions for each fluid. For the cases presented in this paper there are two fluids, and therefore two sets, namely a red and blue fluid denoted by, respectively, $k = R$ and $k = B$, where the red fluid has equal or higher density than the blue fluid. The distribution functions evolve according to the following equation:

\begin{equation}
f_i^k(\vec{x}+\vec{c},t) = f_i^k(\vec{x},t) +  \Omega_i^k ( f_i^k(\vec{x},t))
\label{eq:streamingCG}.
\end{equation}

For the color gradient implementation we use the approach described in \cite{Leclaire2013} and \cite{Leclaire2017}, including the proposed enhanced equilibrium distribution functions. The CG collision operator, $\Omega_i^k $, is a combination of three sub-operators:

\begin{equation}
\Omega_i^k =  (\Omega_i^k)^{(3)} \left[(\Omega_i^k)^{(1)} +(\Omega_i^k)^{(2)} \right]
\label{eq:threeops}.
\end{equation}

 The three separate sub-operators are applied sequentially.

Single phase collision (BGK):
\begin{equation}
f_i^k(\vec{x},t^*) = (\Omega_i^k)^{(1)} f_i^k(\vec{x},t)
\label{eq:BGK}.
\end{equation}

Perturbation step:
\begin{equation}
f_i^k(\vec{x},t^{**}) = (\Omega_i^k)^{(2)} f_i^k(\vec{x},t^*)
\label{eq:perturbation}.
\end{equation}

Recoloring step:
\begin{equation}
f_i^k(\vec{x},t^{***}) = (\Omega_i^k)^{(3)} f_i^k(\vec{x},t^{**})
\label{eq:recoloring}.
\end{equation}

In addition to the collision step, the algorithm includes a streaming step, which is performed in the standard way:
\begin{equation}
f_i^k(\vec{x}+\vec{c}\Delta t,t+\Delta t) = f_i^k(\vec{x},t^{***}) 
\label{eq:streaming}.
\end{equation}

The first sub-operator, $\Omega_i^{k(1)}$, is the original BGK operator as defined in Eq. (\ref{eq:collisSC}) for fluid $k$. 
The second sub-operator, $\Omega_i^{k(2)}$, is known as the perturbation operator. Its function is to apply a force at the interface such that a surface tension is imposed. It takes the form

\begin{equation}
\Omega_i^{k(2)} = \frac{A}{2} |\rho^{N}|  - \left[w_i \frac{(\vec{c}_{i} \cdot \nabla \rho^N)^2}{|\nabla \rho^N|^2} - B_i \right]
\label{eq:perturbationOp}
\end{equation}

with the color field $\rho^N = (\rho_R - \rho_B)/(\rho_R+\rho_B)$ where a value of $\rho^N = -1, 0, 1$ indicate a purely blue fluid, the interface between the red and blue fluid, and a purely red fluid respectively. Based on what is proposed in \cite{Liu} we set $B_0 = -2/9$, $B_{1-6}=1/54$ and $B_{7-18} = 1/27$. By tuning the parameter $A$, the desired surface tension $\sigma$ can be imposed as:

\begin{equation}
\sigma = \frac{4}{9}\frac{A}{\omega_{\text{eff}}}
\label{eq:sigmaA}.
\end{equation}

In Eq. (\ref{eq:sigmaA}) the effective relaxation parameter is set as \cite{Leclaire2017}

\begin{equation}
   \omega_{\text{eff}}  = \frac{2}{6 \overline{\nu} + 1}
\label{eq:effRelax}
\end{equation}

with the kinematic viscosity at the interface, $\overline{\nu}$, being defined by the harmonic density weighted average of $\nu_R$ and $\nu_B$, specifically:

\begin{equation}
    \frac{1}{\overline{\nu}} = \frac{\rho_R}{\rho_R+\rho_B} \frac{1}{\nu_R}    + \frac{\rho_B}{\rho_R+\rho_B} \frac{1}{\nu_B}
\label{eq:interpVisc}.
\end{equation}

The third sub-operator, $\Omega_i^{k(3)}$, is known as the recoloring operator and is used to enforce immiscibility between the fluids. The operator takes the following form for the red and blue fluid:

\begin{equation}
\Omega_i^{R(3)} f_i^R = \frac{\rho_R}{\rho}f_i^* + \beta \frac{\rho_R \rho_B}{\rho^2} \text{cos}(\phi_i) f_i^{eq}|_{\vec{u} = 0}
\label{eq:recoloringOpR},
\end{equation}

\begin{equation}
\Omega_i^{B(3)} f_i^B = \frac{\rho_B}{\rho}f_i^* - \beta \frac{\rho_R \rho_B}{\rho^2} \text{cos}(\phi_i) f_i^{eq}|_{\vec{u} = 0}
\label{eq:recoloringOpB}.
\end{equation}

In Eq. (\ref{eq:recoloringOpR}) and (\ref{eq:recoloringOpB}) $f_i^*$ refers to the total distribution functions as obtained after the perturbation operator, Eq. (\ref{eq:perturbationOp}), has been applied. The total equilibrium distribution is $f_i^{eq} = \sum_k f_i^{k,eq}$ and

\begin{equation}
 \text{cos}(\phi_i) = \frac{\vec{c}_i \cdot \nabla \rho^N}{|\vec{c}_i| |\nabla \rho^N|}
\label{eq:cosphi}.
\end{equation}

The strength of this separation is determined by the tunable parameter $\beta$. For all cases shown in this work, we set $\beta = 0.7$ which is shown to keep the interface as narrow as possible while still resulting in correct interfacial behavior as shown in \cite{Liu}. The density of fluid $k$ is recovered by summing the discrete distributions:
\begin{equation}
\rho_k = \sum_i f_i^k
\label{eq:density}
\end{equation} 

and the total fluid density is given by summing the densities of each fluid:
\begin{equation}
\rho = \sum_k \rho_k
\label{eq:totalDensity}.
\end{equation}

The second moment of the distribution functions corresponds to the total momentum and is calculated as

\begin{equation}
\rho \vec{u}  = \sum_i \sum_k f_i^k \vec{c}_i 
\label{eq:totalMomentum2}.
\end{equation}

The equilibrium distributions entering into Eq. (\ref{eq:BGK}) are now of the following form:

\begin{equation}
f_i^{k,eq}(\rho_k,\vec{u},\alpha_k) = \overline{\nu} \left[ \psi_i (\vec{u} \cdot \nabla \rho_k) + \xi_i (\vec{G}_k : \vec{c}_i \otimes \vec{c}_i  ) \right] 
+ \rho_k \left( \phi_i^k + w_i \left[ 3 \vec{c}_i \cdot \vec{u} + \frac{9}{2} (\vec{c}_i \cdot \vec{u})^2 - \frac{3}{2}(\vec{u} \cdot \vec{u}) \right] \right)
\label{eq:equiliCG}.
\end{equation}

where only for the single phase (BGK) collision step we set $ \phi_i^k = 0 $. In all other cases:

\begin{equation}
   \phi_i^k  = 
    \begin{cases}
      \alpha_k, & i = 1 \\
      (1-\alpha_k)/12 & i = 2 - 7 \\
      (1-\alpha_k)/24 & i = 8 - 19
    \end{cases}
\label{eq:phi}.
\end{equation}

For D3Q19 we use the following set of values reported in \cite{Leclaire2017}: $\psi_0 = -5/2$, $\psi_{1-6}=-1/6$, $\psi_{7-18} = 1/24$ and $\xi_0 = 0$, $\xi_{1-6}=1/4$, $\xi_{7-18} = 1/8$. The tensor $\vec{G}_k$ is defined as 

\begin{equation}
\vec{G}_k = (\vec{u} \otimes \nabla \rho_k ) +  (\vec{u} \otimes \nabla \rho_k )^\intercal 
\label{eq:tensorG}.
\end{equation}

For stability purposes the density ratio between the red and blue fluid should be defined as \cite{Grunau_1993}:

\begin{equation}
   \gamma_{RB}  = \frac{\rho_R^0}{\rho_B^0} = \frac{1-\alpha_B}{1-\alpha_R} 
\label{eq:gamma}
\end{equation}

where $\rho_k^0$ is the initial density of fluid $k$ and we set $\alpha_B = 0.2$. The pressure of each individual fluid of color $k$ is then calculated as

\begin{equation}
   p_{k}  = \rho_k \frac{(1-\alpha_k)}{2} = \rho_k (c_s^k)^2 
\label{eq:CGpressure}
\end{equation}

for the D3Q19 scheme used in this work.

\subsection{Color gradient repulsion force}\label{sec:repforce}
A novel approach to adding a repulsion force to the CG model is presented in this section. A method to accomplish this was previously presented in \cite{Montessori_2019}, where a range-finding algorithm was used to determine the distance between interfaces (e.g. from a neighbouring/impacting droplet). In contrast to that approach, we determine the presence of nearby interfaces by looking for minima and maxima in the $\rho^N$ field. These nodes are located in the exact center between two interfaces. The slight deviation of $\rho^N$ from unity is taken as a measure for the proximity of the interfaces and is used to set an outward (repulsive) force on these nodes. A benefit of this approach is that there is no need to scan outwards from an interface over many grid-nodes, as may be the case in other range-finding algorithms, and only a single layer of boundary-nodes needs to be shared across processes in the case of parallel computing. Hereby no additional overhead or programming is required for parallel computing.

The repulsion force $ \vec{F}_{\text{rep}}$ takes the form

\begin{equation}
 \vec{F}_{\text{rep}}  = \Pi \psi_{\text{rep}}( \rho^N) \sum_i H(-(\vec{n} \cdot \vec{c}_i)(\vec{n}'_i \cdot \vec{c}_i)) \cdot \left(\frac{(\vec{n} \cdot \vec{c}_i)  \vec{n}}{|\vec{c}_i|} +\frac{(\vec{n}'_i \cdot \vec{c}_i)  \vec{n}'_i}{|\vec{c}_i|} \right)
\label{eq:repForce}
\end{equation}

where $\Pi$ is the maximum magnitude of the repulsion force, $H$ represents the Heaviside step function and $\psi_{\text{rep}}(\rho^N)$ determines the drop off rate of the repulsion force with distance $\Delta x$ according to 

\begin{equation}
\psi_{\text{rep}} = (1 + |\rho^N|)^\chi.
\label{eq:potential}
\end{equation}

Specifically the exponent $\chi$ determines the rate at which the repulsion force decays with distance. The exponent is a free parameter and can be tuned according to the desired repulsion force range. The normal $\vec{n}$ at point $\vec{x}$ is calculated as

\begin{equation}
\vec{n}(\vec{x}) = \frac{\nabla \rho^N(\vec{x})}{| \nabla \rho^N(\vec{x}) |} 
\label{eq:normal1}
\end{equation}

and the normals $\vec{n}'_i$ at points $\vec{x}+\vec{c}_i$ are defined as

\begin{equation}
\vec{n}'_i(\vec{x}+\vec{c}_i) = \frac{\nabla \rho^N(\vec{x}+\vec{c}_i)}{| \nabla \rho^N(\vec{x}+\vec{c}_i)|} 
\label{eq:normal2}.
\end{equation}

\section{Simulations}

We will now present the benchmark cases, including setup details, together with a discussion on the purpose of running each case and results. We perform comparative simulations for CG and SC, wherever possible, in order to evaluate the performance of each model for each of the cases presented. It should be noted that in \cite{Leclaire2013} it has been shown that the accuracy of the CG model for dynamic systems can be significantly improved by modifying the equilibrium distribution functions according to the proposed scheme. All cases presented in this paper are simulated using these improved distribution functions, unless otherwise specified. It will be shown in section \ref{ligcont} that for a contracting ligament these improved distribution functions are indeed required to achieve agreement with the analytical solution for the ligament contraction rate.

\subsection{Laplace law test}\label{laplace}
To determine the accuracy of applied surface tension in the CG and SC models we first consider a classical Laplace test, where a static spherical droplet of density $\rho_R$ is initialized in a 3D domain of size $  L_x \times L_y \times L_z =  64 \times 64 \times 64 $ with an ambient fluid density, $\rho_B$. Periodic boundary conditions are applied on all sides of the domain. In subsection \ref{STerror} the numerical surface tension error specific to the CG method is quantified for a range of density- and viscosity-ratios. It is shown that changing the density- or viscosity-ratio has significant effects on the surface tension error. The initial droplet size is also shown to affect the error, where increasing the size of the droplet decreases the surface tension error. In subsection \ref{STCGvsSC} we compare the performance of CG and SC side-by-side. Specifically the accessible parameter range, where the simulations are numerically stable, is considered for both methods as well as the kinetic energy $E_{kin}$, which is a measure of the total spurious currents in the domain.

\subsubsection{Surface tension error trends for CG}\label{STerror}
In all simulations using the CG model the surface tension, $\sigma$, is set a priori as an input parameter, which can be tuned by setting the $A$ parameter in Eq. (\ref{eq:sigmaA}) to the desired value. After allowing the droplet to equilibrate for $10^5$ timesteps, $\sigma$ can be calculated by measuring  $\rho_{in}$ and $\rho_{out}$, the density inside and outside of the droplet, respectively. From these densities the pressures inside and outside of the droplet, $P_{in}$ and $P_{out}$ respectively, are recovered through Eq. (\ref{eq:CGpressure}). Finally the droplet radius $R$ is measured and $\sigma$ is calculated using the Young-Laplace equation:

\begin{equation}
\sigma  = \frac{R \Delta P }{2} 
\label{eq:YoungLaplace}.
\end{equation}

The pressure jump between the inside and the outside of the droplet, $\Delta P = P_{in} - P_{out}$, is measured in the same way as described in \cite{Leclaire2012} by taking 

\begin{equation}
\rho_{in} = \langle \{\rho_R | \rho^N \geq \epsilon \} \rangle
\label{eq:rhoin},
\end{equation}

and

\begin{equation}
\rho_{out} = \langle \{\rho_B | \rho^N \leq -\epsilon \} \rangle 
\label{eq:rhoout}.
\end{equation}

which means that e.g.  $\rho_{in}$ is the average density of $\rho_R$ in the simulation domain, such that $\rho^N$ is greater than or equal to $\epsilon$. We set $\epsilon$ as the value closest to 1 in the set $\{1-0.1^n | n \leq 10 \}$.

To differentiate between the (theoretical) input value and the (calculated) a posteriori value we denote these surface tension values as $\sigma_{th}$ and $\sigma_{cal}$ respectively. Ideally the initially imposed surface tension $\sigma_{th}$ from Eq. (\ref{eq:sigmaA}) coincides with $\sigma_{cal}$ calculated using Eq. (\ref{eq:YoungLaplace}). To quantify the error we define it as $E = |\sigma_{th} - \sigma_{cal}|/\sigma_{th}$ \cite{saito}. It is of interest to know which parameters affect the error and, to this end, a series of simulations is run using varying density- and viscosity-ratios.

\begin{figure}[htp]
\centering
\includegraphics[width=0.75\textwidth]{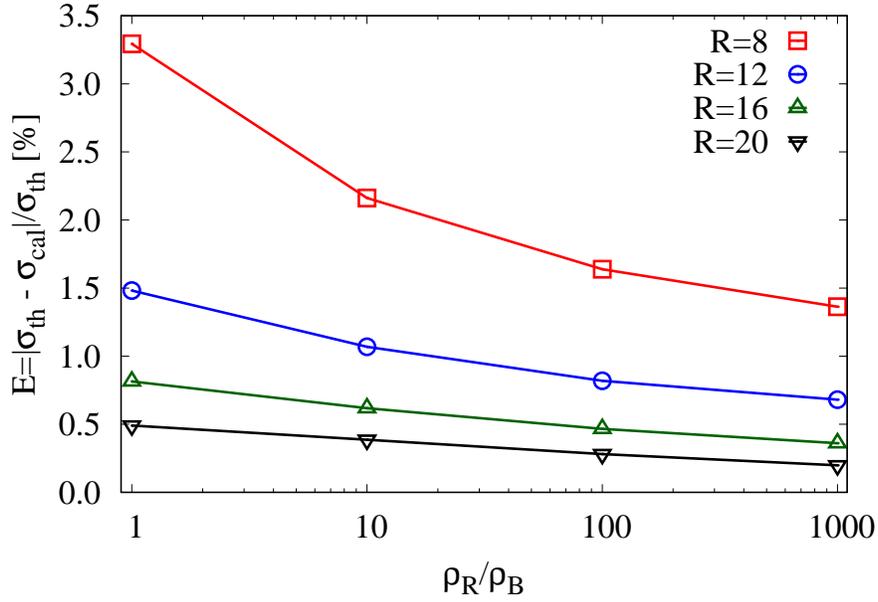}
\caption{\label{fig:vardens} (CG) Surface tension error $E = | \sigma_{th} - \sigma_{cal}|/ \sigma_{th}$  as a function of density ratio, $\rho_R / \rho_B$, for a stationary droplet of density $\rho_R$ surrounded by ambient fluid of density $\rho_B$. Surface tension is kept constant at $\sigma_{th} = 0.01$, with a constant viscosity ratio $\nu_R / \nu_B = 1.0$. The error $E$ decreases both with increasing density ratio and by increasing the droplet radius $R$.}
\end{figure}

\begin{figure}[h!]
\centering
\includegraphics[width=0.75\textwidth]{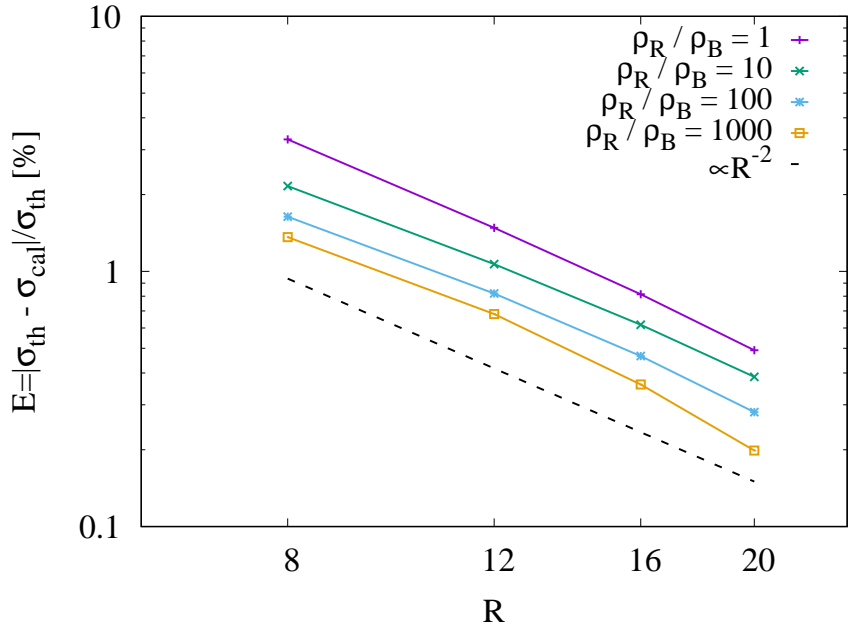}
\caption{\label{fig:varvisc2} (CG) Surface tension error $E = | \sigma_{th} - \sigma_{cal}|/\sigma_{th}$ as a function of $R$ at constant viscosity ratio $\nu_R / \nu_B = 1$ and density ratios $\rho_R / \rho_B = 1, 10, 100, 1000$. Surface tension is kept constant at $\sigma_{th} = 0.01$. The error $E$ decreases with larger droplet radius $R$ with approximately $E \propto R^{-2}$. Exact values of the slope of the fit of each of these sets of values are shown in Table \ref{table:densErrors}.} 
\end{figure}

Two parameter scans are performed with (1) variable density-ratio and constant viscosity-ratio and (2) constant density-ratio and variable viscosity-ratio.
The applied surface tension is kept constant at $\sigma_{th} = 0.01$. The parameter scan is repeated for four different droplet radii, i.e. $R = 8, 12, 16, 20$. The results for the first parameter scan with constant viscosity-ratio $\nu_R/\nu_B = 1$  and varying density-ratio $ \rho_R/\rho_B = 1, 10, 100, 1000$ are shown in Fig. \ref{fig:vardens}. For a given $R$ the error $E$ decreases as the density-ratio increases. The error $E$ also decreases, with increasing $R$ for any density ratio, as approximately $E \propto R^{-2}$, which is shown in Fig. \ref{fig:varvisc2}. Exact values of the slope of the fits on the data shown in  Fig. \ref{fig:varvisc2} are reported in Table \ref{table:densErrors}.

\begin{table}[ht]
\centering
\begin{tabular}[t]{lcccccc}
\hline
$\rho_R / \rho_B $  & 1 & $ 10 $  & 100  & 1000   \\ 
\hline
$b$ &  -2.02 & -1.81  & -1.82  & -1.90   \\ 
\hline
\end{tabular}
\caption{\label{table:densErrors} Fit parameter $b$ is obtained for the fit of the function $E(R) \propto R^b$ using the data presented in Fig. \ref{fig:varvisc2}. Here we report $b$ corresponding to different density ratios.}
\end{table}

\begin{figure}[t!]
\centering
\includegraphics[width=0.75\textwidth]{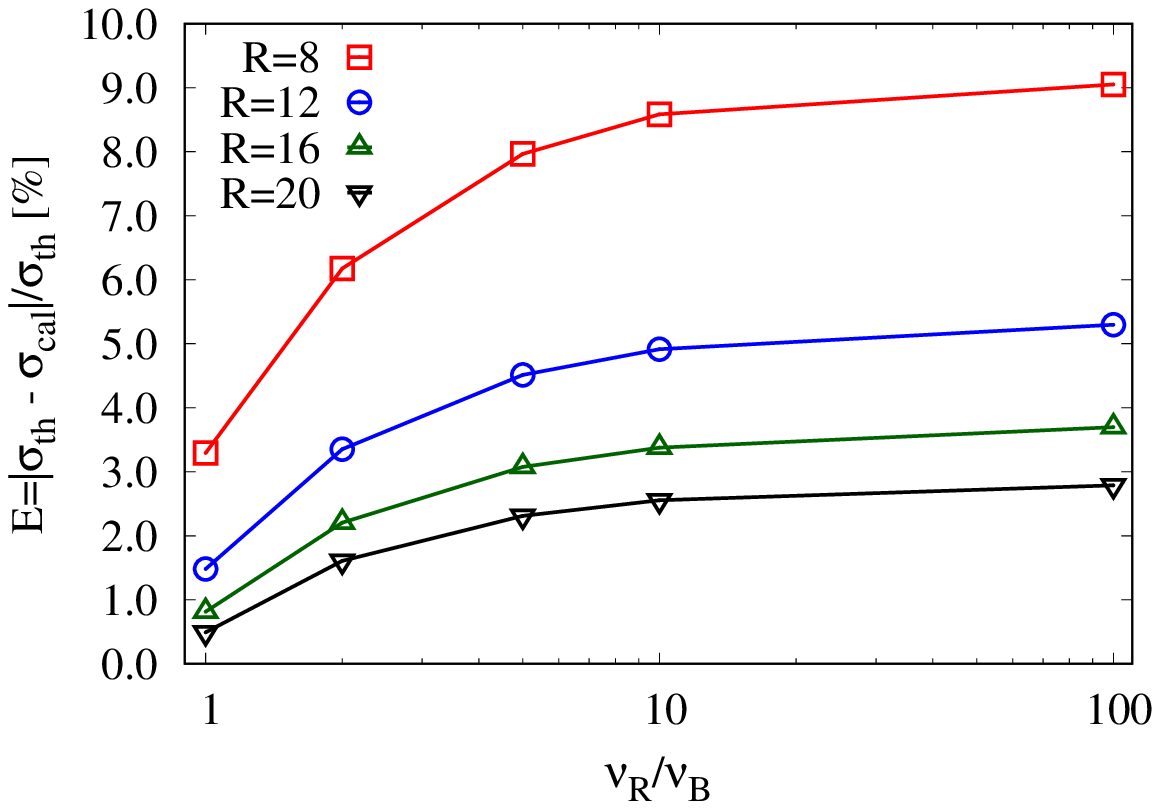}
\caption{\label{fig:varvisc} (CG) Surface tension error $E = | \sigma_{th} - \sigma_{cal}|/ \sigma_{th}$ as a function of viscosity ratio, $\nu_R / \nu_B$, for a stationary droplet of viscosity $\nu_R$ surrounded by ambient fluid of viscosity  $\nu_B$. Surface tension is kept constant at $\sigma_{th} = 0.01$, with a constant density ratio $\rho_R / \rho_B = 1.0$. The error $E$ increases with increasing viscosity ratio but decreases with larger droplet radius $R$.} 
\end{figure}

\begin{figure}[h!]
\centering
\includegraphics[width=0.75\textwidth]{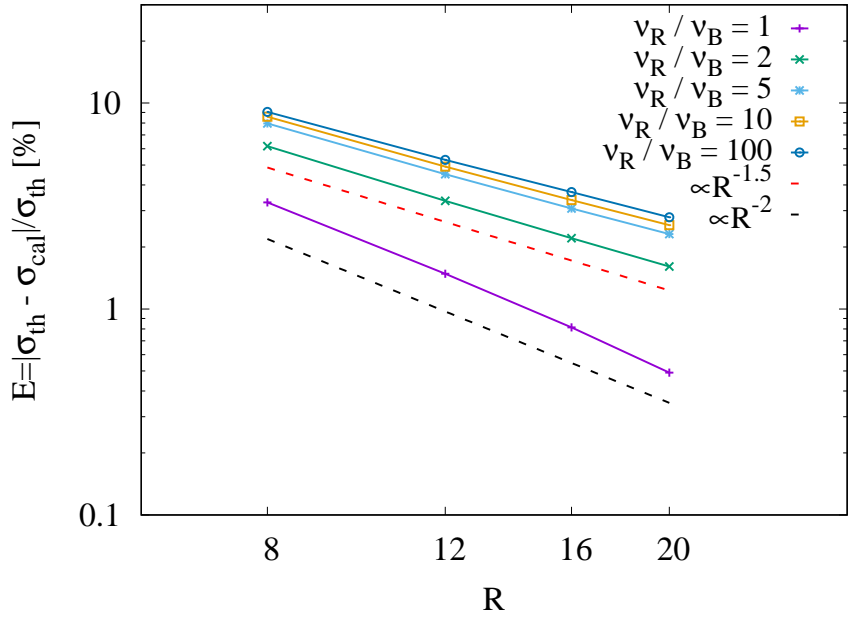}
\caption{\label{fig:varvisc3} (CG) Surface tension error $E = | \sigma_{th} - \sigma_{cal}|/ \sigma_{th}$ as a function of $R$ at viscosity ratios $\nu_R / \nu_B =1, 2, 5, 10, 100$ and  constant density ratio $\rho_R / \rho_B = 1$. Surface tension is kept constant at $\sigma_{th} = 0.01$. The error $E$ decreases with larger droplet radius $R$ with $E \propto R^{-2}$ for $\nu_R / \nu_B = 100$ and with $E \propto	 R^{-1.5}$ for all other viscosity ratios. Exact values of the slope of the fit of each of these sets of values are shown in Table \ref{table:viscErrors}.} 
\end{figure}

Next we consider the second parameter scan with a constant density-ratio $\rho_R/\rho_B = 1$ and a variable viscosity-ratio $\nu_R/\nu_B = 1, 2, 5, 10, 100$, the results for which are reported in Fig. \ref{fig:varvisc}. In this case we see that the error increases as the viscosity ratio increases. However, increasing $R$ still leads to a reduced error for all viscosity-ratios as illustrated in Fig. \ref{fig:varvisc3} where $E$ is plotted as a function of $R$. For the case with the highest viscosity ratio, $\nu_R/\nu_B = 100$, we find approximately $E \propto R^{-2}$ and for all lower viscosity ratios we find, approximately $E \propto R^{-1.5}$. Exact values for the slopes of the fits on the data shown in  Fig. \ref{fig:varvisc3} are reported in Table \ref{table:viscErrors}. We can conclude that increasing the density ratio reduces the surface tension error, whereas increasing the viscosity ratio will increase the error. In both cases however the error is significantly reduced by increasing the droplet radius.

\begin{figure}[h!]
\centering
\includegraphics[width=0.75\textwidth]{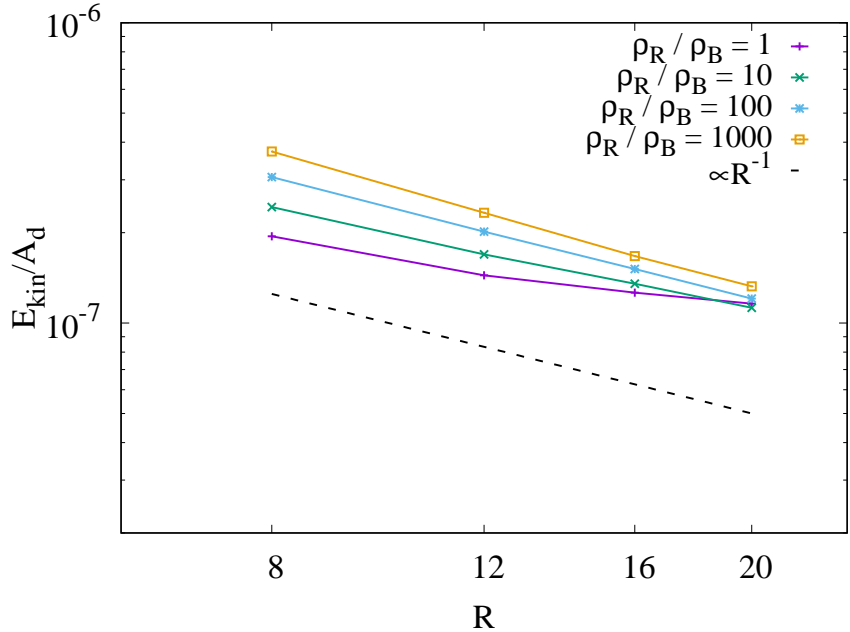}
\caption{\label{fig:radiusErrors} (CG) The total kinetic energy, $E_{kin}$ (integrated over the full simulation domain) normalized by the droplet surface area $A_d=4 \pi R^2$ as a function of radius $R$. Radius $R = 8, 12, 16, 20$ at constant viscosity ratio $\nu_R / \nu_B = 1$ and density ratios $\rho_R / \rho_B = 1, 10, 100, 1000$. Surface tension is kept constant at $\sigma_{th} = 0.01$. The decrease in normalized kinetic energy, $E_{kin} / A_d$, is approximately proportional to $E_{kin} / A_d \propto R^{-1}$  except for the $\rho_R / \rho_B = 1$ case, which deviates from the trend significantly when $R >12$. Exact values of the slope of the fit of each of these sets of values are shown in Table \ref{table:radiusErrors}.}
\end{figure}

Finally, we investigate the effect of droplet radius $R$ on the spurious currents and thereby the total kinetic energy $E_{kin}$ in the domain. Specifically, the total kinetic energy, $E_{kin}$ (integrated over the full simulation domain) is normalized by the droplet surface area $A_d=4 \pi R^2$ as a function of radius $R$. Radius $R = 8, 12, 16, 20$ at constant viscosity ratio $\nu_R / \nu_B = 1$ and density ratios $\rho_R / \rho_B = 1, 10, 100, 1000$. Surface tension is kept constant at $\sigma_{th} = 0.01$. The results are reported in Fig. \ref{fig:radiusErrors} and we find that the decrease in normalized kinetic energy, $E_{kin} / A_d$, is approximately proportional to $E_{kin} / A_d \propto R^{-1}$. A notable exception is the $\rho_R / \rho_B = 1$ case, which deviates from the trend significantly when $R >12$. Exact values of the slope of the fit of each of these sets of values are shown in Table \ref{table:radiusErrors}. From this we can conclude that, generally, lower curvature of the droplet, i.e. larger $R$, decreases $E_{kin}$ per unit of surface area $A_d$.

\begin{table}[ht]
\centering
\begin{tabular}[t]{lccccccc}
\hline
$\nu_R / \nu_B $ &  1 & $ 2 $ & 5  & 10 & 100   \\ 
\hline
$b$ &  -2.02 & -1.48 & -1.37  & -1.34  &  -1.29 \\ 
\hline
\end{tabular}
\caption{\label{table:viscErrors} Fit parameter $b$ is obtained for the fit of the function $E(R) \propto R^b$ using the data presented in Fig. \ref{fig:varvisc3}. Here we report $b$ corresponding to different viscosity ratios.}
\end{table}

\begin{table}[ht]
\centering
\begin{tabular}[t]{lcccccc}
\hline
$\rho_R / \rho_B $  & 1 & $ 10 $  & 100  & 1000   \\ 
\hline
$b$ &  -0.59 & -0.85 & -1.02  & -1.14   \\ 
\hline
\end{tabular}
\caption{\label{table:radiusErrors} Fit parameter $b$ is obtained for the fit of the function $E_{kin}/A_d \propto R^b$ using the data presented in Fig. \ref{fig:radiusErrors}. Here we report $b$ corresponding to different density ratios.}
\end{table}

\subsubsection{Comparison of CG to SC}\label{STCGvsSC}

\begin{figure}[htp]
\centering
\includegraphics[width=1.0\textwidth]{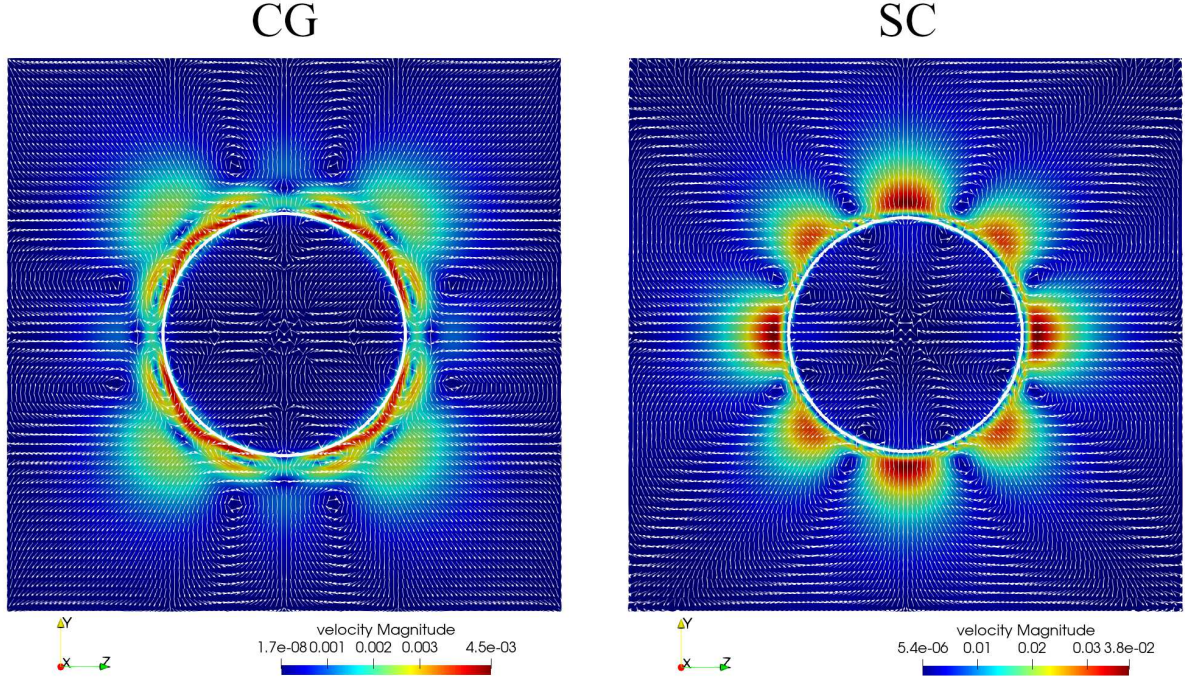}
\caption{\label{fig:CGandSCspuriousCurrents} (CG vs. SC) Cross-section of the velocity field of a (3D) stationary droplet simulation with a vector plot overlay indicating the direction of the spurious currents. The droplet surface is indicated by the white line. The input parameters used are reported in Table \ref{table:dropOscParams}. For the CG case (pictured on the left) the maximum spurious currents are both less strong (by nearly an order of magnitude) and are closer to the interface, compared to tthe SC case (pictured on the right).}
\end{figure}

By running identically initialized simulations for the CG and SC model, it is possible to compare their performance side-by-side. For the initialization all relevant parameters are matched, i.e. $R$, $\rho_R$, $\rho_B$, $\nu_R$, $\nu_B$  and $\sigma$. For the SC method the coupling parameter $G$ determines $\sigma$ and also the equilibrium values of $\rho_R$ and $\rho_B$, meaning that the surface tension and the densities cannot be set independently from each other. This is in contrast with the CG model where these parameters are not coupled. Due to this interdependency of variables in the SC model it is necessary to first determine $\sigma$ and equilibrium $\rho_R$ and $\rho_B$ as a function of $G$, after which the parameters can be matched in a comparative CG simulation. For both methods we will then explore: (1) the accessible parameter range in terms of density ratios and (2) the kinetic energy $E_{kin}$ over the entire domain, which we take as a measure of the intensity of spurious currents. Since we are considering a stationary system composed of a quiescent droplet in ambient fluid, ideally $E_{kin} = 0$. The presence of (unphysical) spurious currents - mainly close to the droplet-ambient interface - is a common feature of most multiphase schemes and, depending on the magnitude, it may influence the results of the simulations. To illustrate the phenomenon of spurious currents we took a cross-section of the velocity field of a 3D simulation, shown in Fig. \ref{fig:CGandSCspuriousCurrents} for both the CG and SC cases. The input parameters used are the same as for the droplet oscillation case, as reported in Table \ref{table:dropOscParams}. Note that the areas with the most significant spurious currents are closer to the interface in the CG simulation, compared to the SC simulation. Furthermore we see that the maximum spurious current in the CG simulation is an order of magnitude smaller.

\begin{figure}[h!]
\centering
\includegraphics[width=0.75\textwidth]{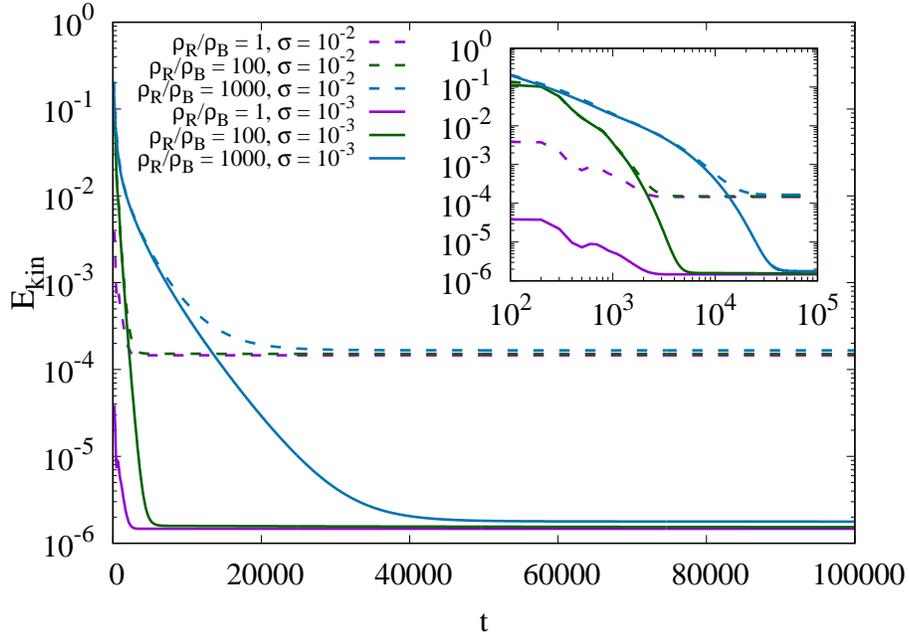}
\caption{\label{fig:EkinLaplaceInTimeAll} (CG) The total kinetic energy, $E_{kin}$, integrated over the full simulation domain as a function of time, $t$, for a stationary 3D droplet with initial radius $R = 20$. The stabilization of $E_{kin}$ takes longer for higher values of  $\rho_R / \rho_B$ and for lower values of $\sigma$. In all cases $10^5$ timesteps was found to be sufficient for the system to fully equilibrate for any combination of $\rho_R / \rho_B$ and $\sigma$ considered in this work. The inset graph shows the same data with logarithmic scales on both the horizontal and vertical axes.}
\end{figure}

\begin{figure}[h!]
\centering
\includegraphics[width=0.75\textwidth]{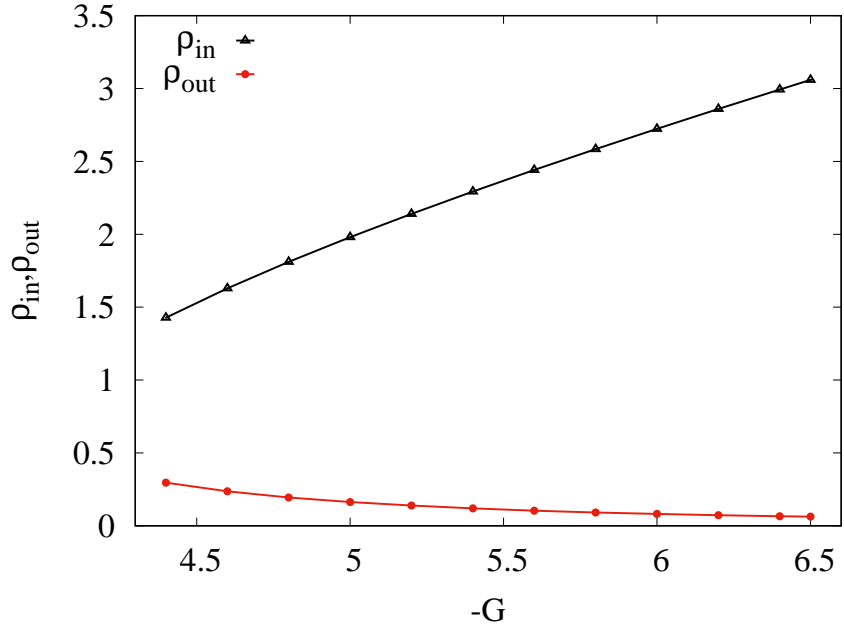}
\caption{\label{fig:GvsRho} (SC) Droplet density, $ \rho_{in}$, and ambient density, $ \rho_{out} $, after allowing a stationary droplet to fully equilibrate as a function of coupling parameter $G$ which, for the Shan-Chen pseudopotential method, dictates both the density ratio and surface tension.}
\end{figure}

\begin{figure}[h!]
\centering
\includegraphics[width=0.75\textwidth]{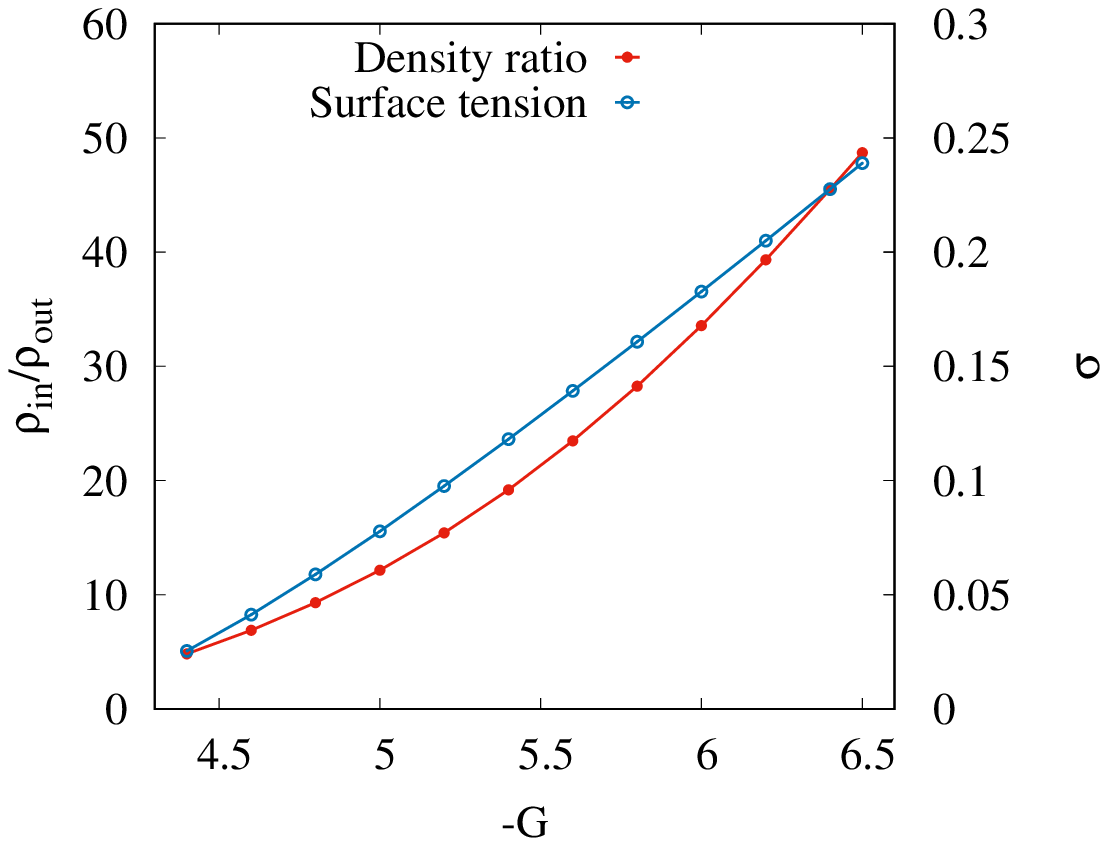}
\caption{\label{fig:GvsRhoRatio} (SC) Density ratio $ \rho_{in}  /   \rho_{out} $ (red filled circles) and associated surface tension $\sigma$ (light blue hollow circles) after allowing a stationary droplet to fully equilibrate as a function of the coupling parameter $G$ which, for the Shan-Chen pseudopotential method, dictates both the density ratio and the surface tension. The maximum density ratio attainable, while maintaining numerical stability, is approximately 50.}
\end{figure}

For all simulations presented in this section the droplet is initialized with a radius $R = 20$ and allowed to equilibrate for $10^5$ timesteps, after which $E_{kin}$ has stabilized, see Fig. \ref{fig:EkinLaplaceInTimeAll}. The time for stabilization depends on $\rho_R / \rho_B$ and $\sigma$, a higher density ratio and a lower surface tension value require a longer equilibration time, however for all cases considered in this work stability was achieved after $10^5$ timesteps. As mentioned, for the SC model $\sigma$ is determined by the coupling parameter $G$. A numerically stable parameter range is found for $ 4.2 < G < 6.5 $, outside of which the simulations were found to be numerically unstable. The associated equilibrium densities, $\rho_{in}$ and $\rho_{out}$, are shown in Fig. \ref{fig:GvsRho} and the associated density ratios and measured $\sigma$, as a function of $G$, are shown in Fig. \ref{fig:GvsRhoRatio}. Numerically stable simulations for this setup were limited to a maximum of approximately $\rho_R / \rho_B = 50$. The surface tension $\sigma$ was determined at the end of each simulation through the Young-Laplace equation, Eq. (\ref{eq:YoungLaplace}).

\begin{figure}[h!]
\centering
\includegraphics[width=0.75\textwidth]{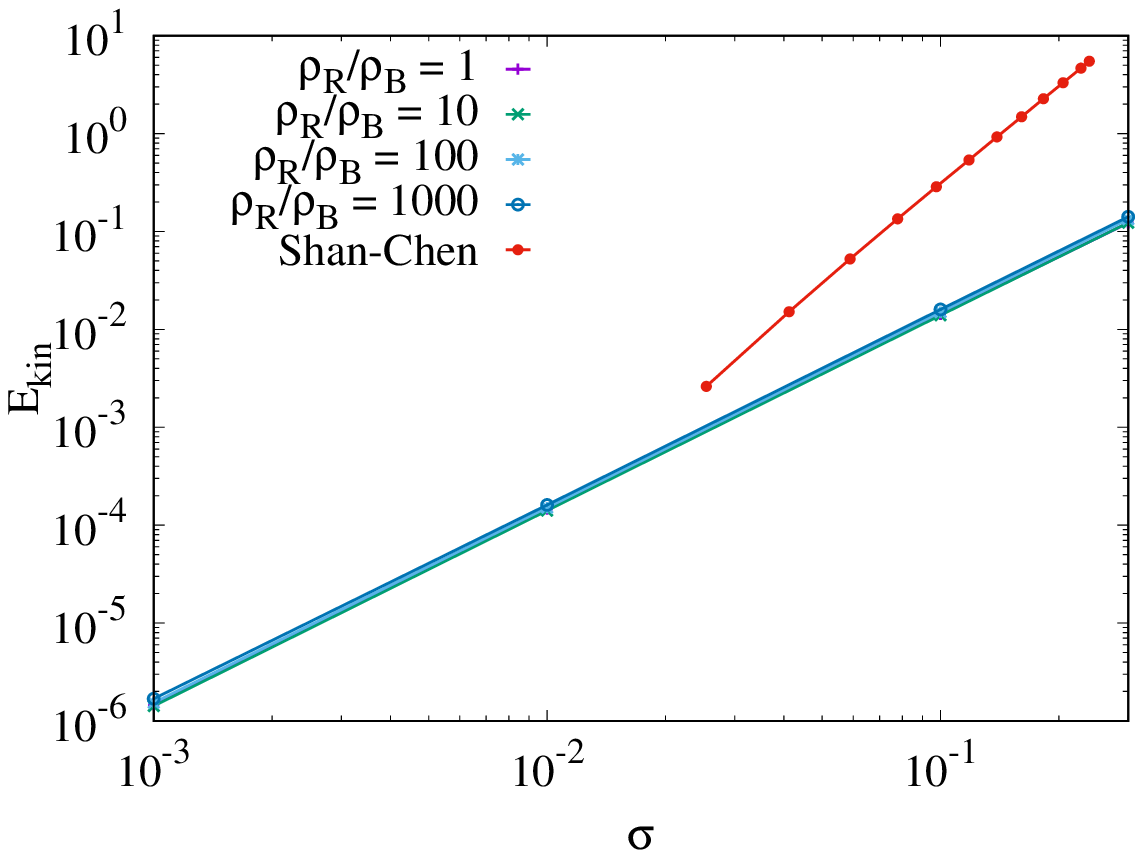}
\caption{\label{fig:sigmaVsEkinCompareSC} (CG vs. SC) The total kinetic energy, $E_{kin}$, integrated over the simulation domain as a function of the surface tension, $\sigma$, for a stationary 3D droplet with initial radius $R = 20$. The SC method is compared to the CG method where a range of simulations is run with varying density ratios $\rho_R / \rho_B = 1, 10, 100, 1000$. Note that for the SC method the density ratio is linked to the surface tension and both parameters cannot be varied independently. The CG method shows a much wider stable parameter range in terms of $\sigma$ and lower $E_{kin}$ overall. In both cases $E_{kin}$ increases with increasing $\sigma$, albeit at a slower rate in the CG case. By fitting the data we find $E_{kin} \propto \sigma^{2}$  for CG and $E_{kin} \propto \sigma^{3.3}$ for SC.} 
\end{figure}

As a measure of spurious currents, $E_{kin}$ is determined and found to increase by approximately 3 orders of magnitude over the considered surface tension range for the SC case, as shown in Fig. \ref{fig:sigmaVsEkinCompareSC}. Simulations are also run with the CG model to cover the same range of $\sigma$ and beyond. Since the density ratio and $\sigma$ are uncoupled for CG this allows us to explore the effect of different density ratios on $E_{kin}$. To this end the simulations are repeated for $\rho_R / \rho_B = 1, 10, 100, 1000$. The results are shown in Fig. \ref{fig:sigmaVsEkinCompareSC}.  It is clear that for both methods there is an increase of $E_{kin}$ when $\sigma$ is increased and there is near perfect overlap for different density ratios indicating a very weak dependence of $E_{kin}$ on the density ratio. Not only is a much wider range of $\sigma$ accessible when using CG, but also in all cases $E_{kin}$ is significantly lower than for the identical case using SC. In the most extreme case (with the highest $\sigma$) there is nearly two orders of magnitude difference for $E_{kin}$ between the CG and SC simulations, see Fig. \ref{fig:sigmaVsEkinCompareSC}. We found that for $\sigma > 0.3$ the simulations became numerically unstable. Although setting $\sigma < 10^{-3}$ was numerically stable, the increased equilibration time required (as shown in Fig. \ref{fig:EkinLaplaceInTimeAll}) leads to impractically long simulation times and were therefore omitted in this work. 

\begin{figure}[h!]
\centering
\includegraphics[width=0.75\textwidth]{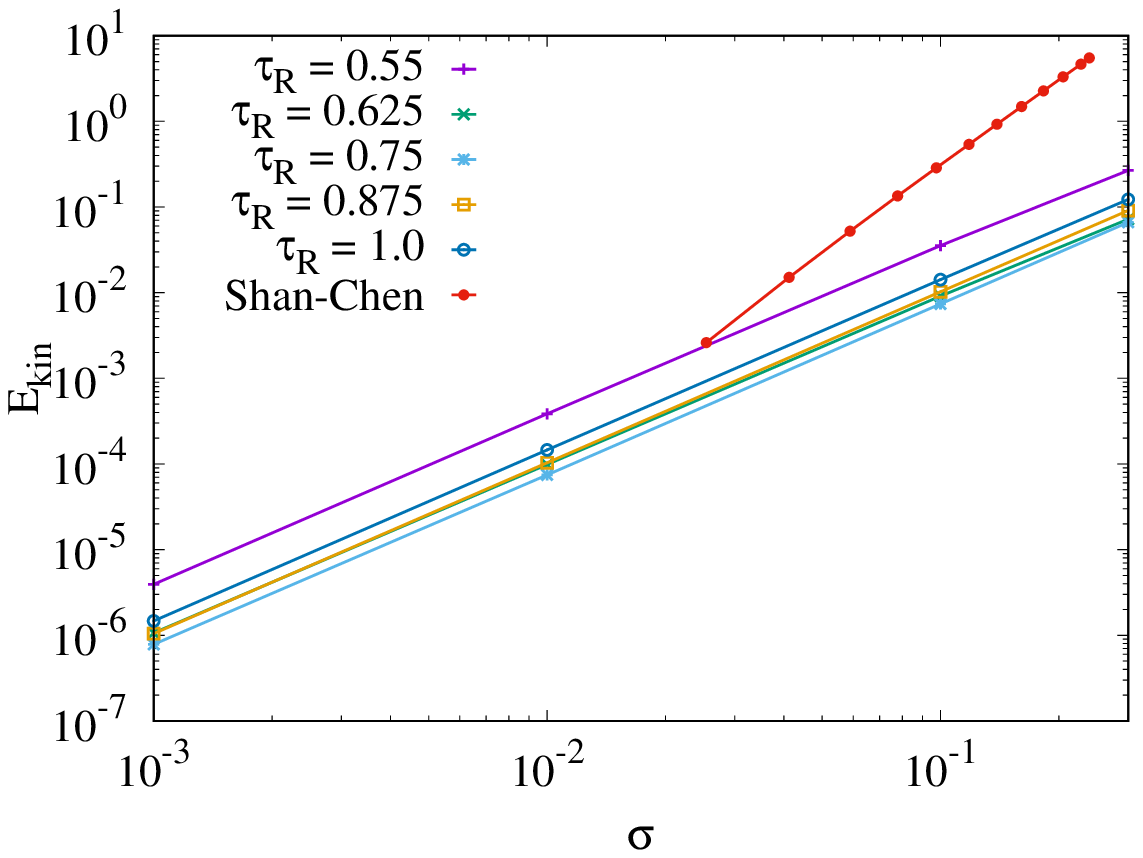}
\caption{\label{fig:sigmaVsEkinCompareSCvisc} (CG vs. SC) The total kinetic energy, $E_{kin}$, integrated over the simulation domain as a function of the surface tension $\sigma$ for a stationary 3D droplet with initial radius $R = 20$. The SC method is compared to the CG method where a range of simulations is run with varying viscosities by changing the value of the relaxation time, $\tau_R$, while the density ratio is kept constant at $\rho_R / \rho_B = 1$. Note that for the SC method the density ratio is linked to the surface tension and both parameters cannot be varied independently. The CG method shows a much wider stable parameter range in terms of $\sigma$ and lower $E_{kin}$ overall. In both cases $E_{kin}$ increases with increasing $\sigma$, albeit at a slower rate in the CG case. By fitting the data we find $E_{kin} \propto \sigma^{2}$  for CG and $E_{kin} \propto \sigma^{3.3}$ for SC.} 
\end{figure}

We now consider also the effect of changing the relaxation time, $\tau_R$, which sets the red fluid viscosity in the CG method. For simplicity we set $\rho_R / \rho_B = 1$ and $\tau_B = 1$. As it can be seen from the results reported in Fig. \ref{fig:sigmaVsEkinCompareSCvisc}, $E_{kin}$ changes by at most 1 order of magnitude within the range of $\tau_R = 0.55, 0.625, 0.75, 0.875, 1.0$ for any investigated $\sigma$. A trend is not immediately obvious however, as $E_{kin}$ is highest for $\tau_R = 0.55$ and lowest for $\tau_R = 0.75$ with $\tau_R = 1$ being in between.

We can conclude that the SC model has limited numerical stability for higher density ratios. A maximum of approximately $\rho_R / \rho_B = 50$ was found. This is  too low for simulating an ink-air system with realistic parameters where usually $\rho_R / \rho_B \approx 1000$, which is achievable with the CG model. An impractical aspect of the SC model is the limited accessible parameter space, due to the fact that the density and surface tension are coupled. The CG model does not suffer from this drawback. Finally the total spurious currents in the domain - as measured through $E_{kin}$ - is lower for every case considered in this section when using the CG model.

\subsection{Droplet oscillation}\label{droposc}

The Laplace law test from section \ref{laplace} was used to quantify numerical errors in imposing the desired surface tension and explore accessible parameter ranges for both the CG and SC model in a static system. To gain information on physical accuracy in a dynamic system we now perform a series of droplet oscillation simulations. In this case an ellipsoidal droplet is initialized and it will contract into a spherical droplet after several oscillations due to the effect of surface tension. The exact simulation parameters used are reported in Table \ref{table:dropOscParams}. 

\begin{table}[ht]
\centering
\caption{\label{table:dropOscParams} Input parameters for the droplet oscillation simulations: red and blue fluid density $\rho_R$ and $\rho_B$, relaxation times $\tau_R$ and $\tau_B$, surface tension $\sigma$ and coupling parameter $G$.}
\begin{tabular}[t]{lcccccccc}
\hline
Set & Method & $\rho_R$  & $\rho_B$   & $\tau_R$    & $\tau_B$   & $\sigma$  & $G$   \\ 
\hline
1 & CG     & 2.73 &  0.073 & 1.0  & 1.0  & 0.183 & - \\ 
2 & SC     & 2.73 &  0.073 & 1.0  & 1.0 & - & $-6$ \\ 
\hline
\end{tabular}
\end{table}

\begin{figure}[h]
\centering
\includegraphics[width=0.75\textwidth]{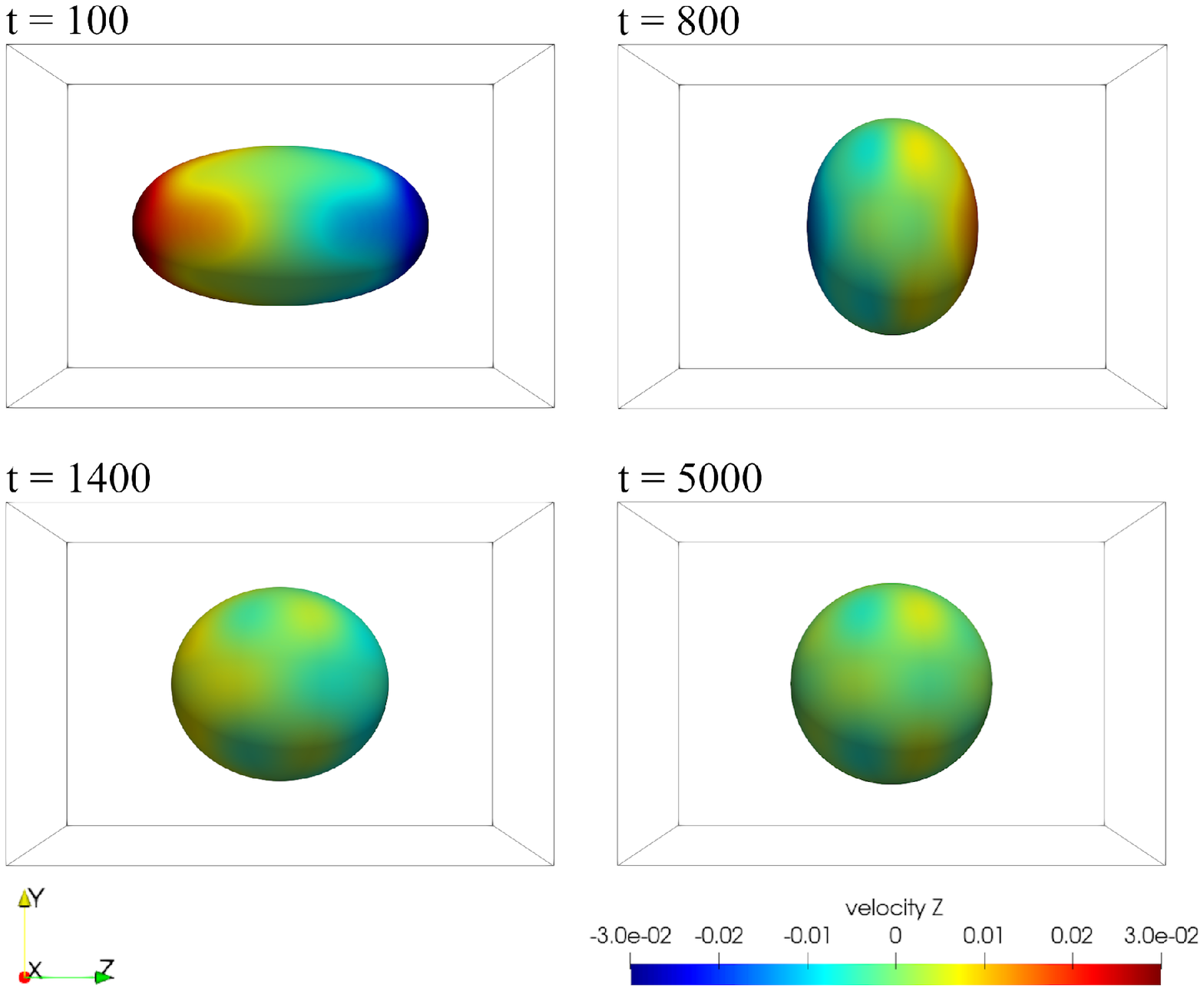}
\caption{\label{fig:elipseVisualization} (CG) Visualization of the interface of a typical oscillating droplet simulation at times $t = 100, 800, 1400, 5000$ with $\sigma = 0.183$, $\rho_R = 2.73$, $\rho_B = 0.073$, $\tau_R = \tau_B = 1.0$. The final frame shows a droplet at rest with equilibrium radius $R_D = 24.965$. The isosurface is drawn at $\rho^N = 0$ and colored by the z-component of the velocity, with the z-axis being along the length of the domain. The non-zero velocities visible for the droplet at rest (e.g. at $t = 5000$) are due to spurious currents.}
\end{figure}

\begin{figure}[h]
\centering
\includegraphics[width=0.75\textwidth]{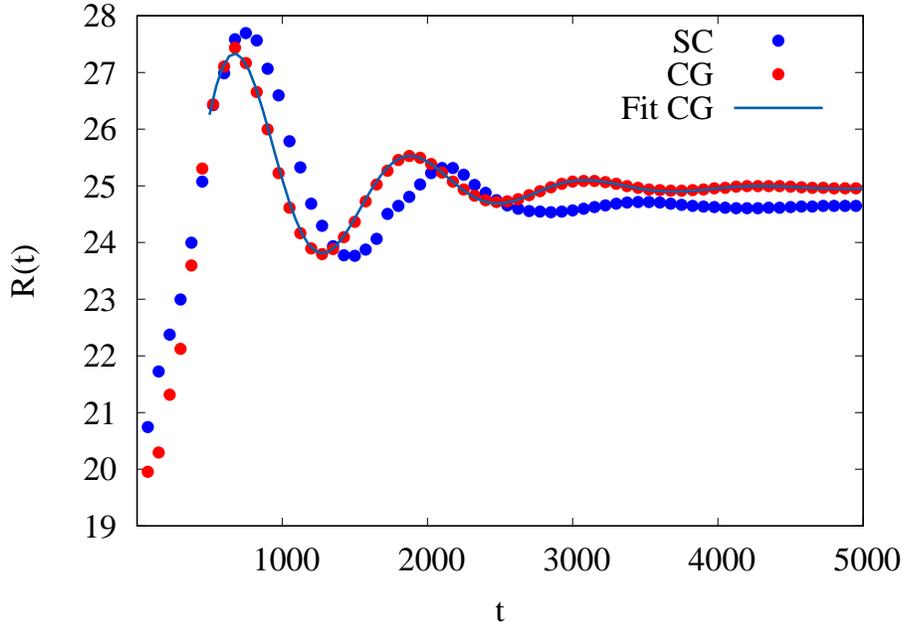}
\caption{\label{fig:diameterInTime} (CG vs. SC) Radius as a function of time $R(t)$  of an oscillating droplet as measured along the y-axis in the center of the domain with equilibrium radius $R_D = 24.965$ for CG (red dots) and $R_D = 24.625$ for SC (blue dots). Parameters are set as $\sigma = 0.183$, $\rho_R = 2.73$, $\rho_B = 0.073$, $\tau_R = \tau_B = 1.0$. The droplet (initialized as an ellipsoid) oscillates and a fit is made of the simulation data such that the oscillation frequency $\omega_2$ can be determined through the generated fit.}
\end{figure}

A visualization illustrating the oscillatory behaviour is shown in Fig. \ref{fig:elipseVisualization}. The oscillations preceding the final steady state occur at a certain frequency which will be compared to the expected theoretical oscillation frequency of the second mode $\omega_2$ for which the analytical solution is reported in \cite{Miller1968} as

\begin{equation}
 \omega_2 = \omega_2^* - 0.5 \alpha \sqrt{\omega_2^*} + 0.25 \alpha^2
\label{eq:millscriv}
\end{equation}

with 

\begin{equation}
\omega_2^* = \sqrt{\frac{24\sigma}{R_D^3 (2 \rho_{out} + 3 \rho_{in})}} 
\label{eq:millscriv2}
\end{equation}

and where the parameter $\alpha$ is defined as

\begin{equation}
\alpha = \frac{25\sqrt{\nu_{in}\nu_{out}} \rho_{in} \rho_{out} }{\sqrt{2} R_D (2 \rho_{out} + 3 \rho_{in} ) (\sqrt{\nu_{in}} \rho_{in} + \sqrt{\nu_{out}} \rho_{out})}
\label{eq:millscriv3}
\end{equation}

with the equilibrium radius, $R_D$, density inside and outside the droplet $\rho_{in}$ and $\rho_{out}$, respectively, and kinematic viscosity inside and outside the droplet $\nu_{in}$ and $\nu_{out}$, respectively. To obtain the oscillation frequency from the simulation, the droplet radius $R(t)$ is tracked as a function of time and a fit is created for the simulation data, as illustrated in Fig. \ref{fig:diameterInTime}. $R(t)$ is measured across the y-axis (in Fig. \ref{fig:elipseVisualization}) at the center-point of the domain i.e. at $x = L_x / 2$ and $z = L_z /2$. For this case we define the interface to be at the point where $\rho = (\rho_R^0 - \rho_B^0)/2$ for the SC case and at $\rho^N = 0$ for the CG case. To get more accurate data (on a sub-grid scale) a linear interpolation is performed between the two adjacent data-points closest to the interface. From the resulting fit, using the equation $f(t) = R_D + a \exp(-b\cdot t) \sin(d \cdot t+c)$, the oscillation frequency $\omega_2$ (corresponding to the fit parameter $d$) is determined. This is repeated for four different droplet sizes with approximate equilibrium radii (i.e. the radii of the steady state spherical droplets) $R_D = 12, 25, 37, 49$. It should be noted that the exact value of $R_D$ can differ between CG and SC simulations slightly. This is mostly due to initialization effects, since the ellipsoid is initialized with a sharp interface, after which the interface equilibrates and spreads out. The interface thickness differs slightly between the CG and SC models. Also some numerical evaporation may occur in the SC case, which is not the case with the CG model due to the phases being forcefully separated by the recoloring collision operator.  Exact values of the fit for $R_D$, $\omega_2$ and $E$ are reported in Table \ref{table:dropOsc}.

\begin{table}[ht]
\centering
\caption{\label{table:dropOsc} Measured values for droplet oscillation simulations: equilibrium radius $R_D$, oscillation frequency $\omega_2$ and error $E$.}
\begin{tabular}[t]{lcccc}
\hline
Method & $R_D$  & $\omega_2$          & $E [\%]$         \\ 
\hline
CG     & 11.65 & $14.65 \cdot 10^{-3}$  & 11.51     \\ 
       & 24.97 & $5.20 \cdot 10^{-3}$ & 2.49   \\ 
       & 36.97 & $2.97 \cdot 10^{-3}$ & 0.31       \\ 
       & 48.95 & $1.96 \cdot 10^{-3}$ & 0.16       \\ 
\hline
SC     & 11.79 & $1.17 \cdot 10^{-3}$ & 27.96   \\ 
       & 24.62 & $4.44 \cdot 10^{-3}$ & 15.45       \\ 
       & 37.02 & $2.62 \cdot 10^{-3}$ & 10.52        \\ 
       & 48.60 & $1.94 \cdot 10^{-3}$ & 7.77        \\ 
\hline
\end{tabular}
\end{table}

\begin{figure}[h]
\centering
\includegraphics[width=0.75\textwidth]{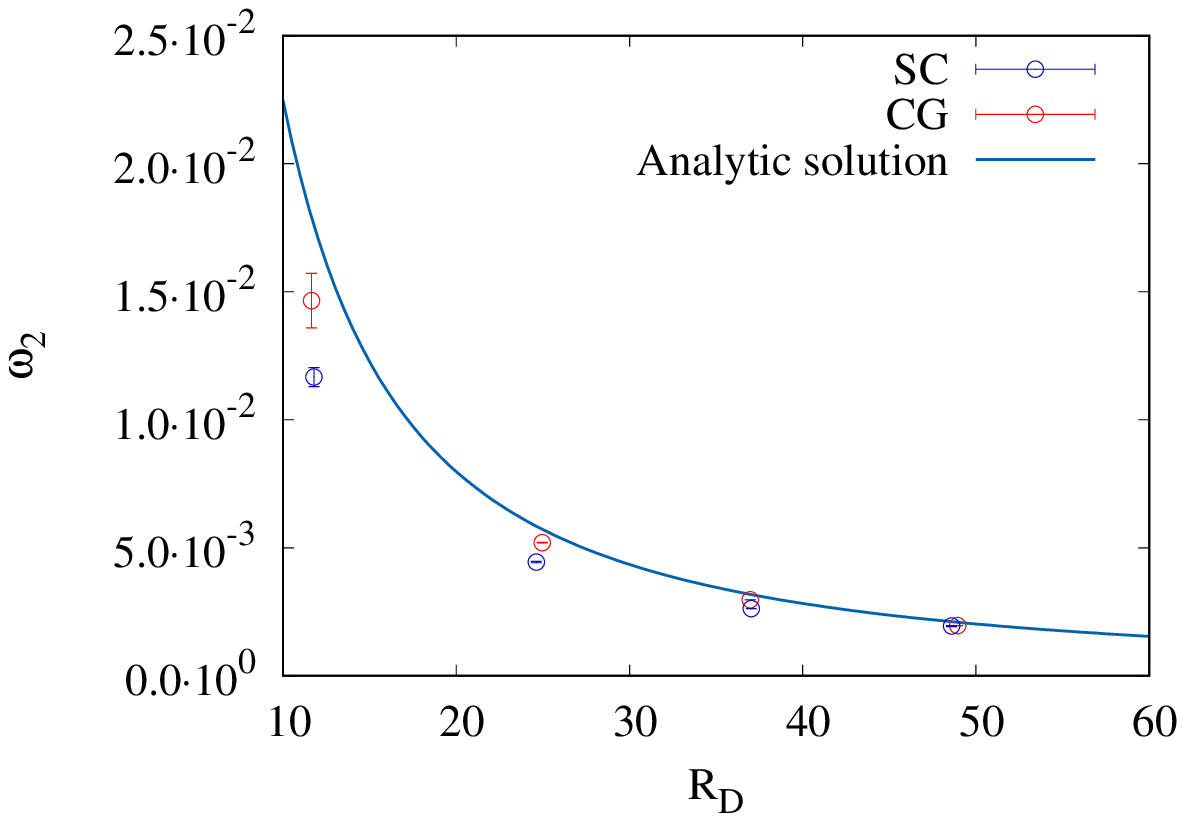}
\caption{\label{fig:frequencyVsRadiusSCvsCG}  (CG vs. SC) Oscillation frequency $\omega_2$ as a function of four different equilibrium radii $R_D = 12, 25, 37, 49$. Initial parameters are $\sigma = 0.183$, $\rho_R = 2.73$, $\rho_B = 0.073$, $\tau_R = \tau_B = 1.0$. The simulation data for CG is in good agreement with the analytical solution with the difference being particularly small for higher values of $R_D$. The error bars represent the asymptotic standard error associated with the fit parameter $d$, corresponding to the oscillation frequency $\omega_2$.}
\end{figure}

\begin{figure}[h]
\centering
\includegraphics[width=0.75\textwidth]{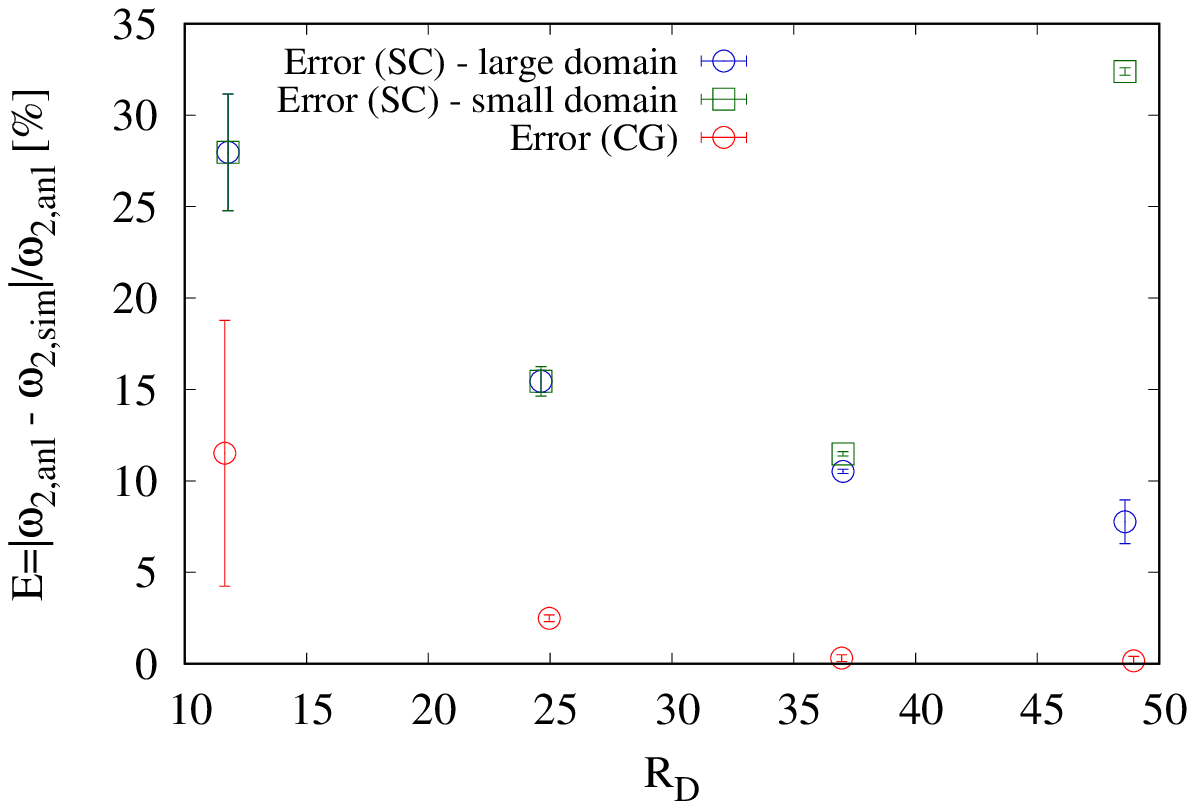}
\caption{\label{fig:frequencyVsErrorSCvsCG} (CG vs. SC) Error of the droplet oscillation frequency  $E=|\omega_{2,anl} - \omega_{2,sim}|/\omega_{2,anl}$ as a function of equilibrium drop radius $R_D$. For the CG case the error $E$ decreases monotonically with increasing $R_D$. For the comparable SC cases there is a similar trend, as long as the domain size is sufficiently large with respect to $R_D$. Increasing the domain size from $L_x \times L_y \times L_z = 160 \times 160 \times 240$ to $L_x \times L_y \times L_z =  240 \times 240 \times 360$ significantly reduces the error for SC at the highest $R_D$. The error bars represent the asymptotic standard error associated with the fit parameter $d$, corresponding to the oscillation frequency $\omega_2$.}
\end{figure}

The measured frequency $\omega_2$ as a function of $R_D$ is plotted in Fig. \ref{fig:frequencyVsRadiusSCvsCG} where the analytic solution is plotted alongside the simulation results for both CG and SC simulations. We define the error with respect to the analytical solution as $E = |\omega_{2,anl} - \omega_{2,sim}| / \omega_{2,anl}$, and $E$ as a function of $R_D$ is plotted in Fig. \ref{fig:frequencyVsErrorSCvsCG} and shows significantly higher errors for all SC simulations with $7.77\% < E < 27.96\%$. In the CG cases $E$ is smaller for all values of $R_D$. The CG simulations error ranges from $0.16\% < E < 11.51\%$ . A resolution effect is seen where a higher $R_D$ in general gives a lower $E$. In particular for $R_D = 37, 49$ in the CG case the agreement with the analytical solution is very close, i.e. $E < 1 \%$. For the smallest droplet considered ($ R_D = 12 $) the error is significantly higher at approximately $11.51\%$. For the SC model the errors are generally larger, but also decrease monotonically with increasing $R_D$ as long as the domain size is sufficiently large. It was found that for the domain size $L_x \times L_y \times L_z = 160 \times 160 \times 240$ the error increases drastically for $R_D = 49$ due to the effect of spurious currents generated on one side of the droplet interacting with the opposite side of the droplet (which is possible due to periodic boundary conditions). By increasing the domain size to $L_x \times L_y \times L_z =  240 \times 240 \times 360$ the error once again follows the expected trend. The errors for the fits, from which the oscillation frequency is obtained, are shown as error bars in Fig. \ref{fig:frequencyVsRadiusSCvsCG} and Fig. \ref{fig:frequencyVsErrorSCvsCG}. Specifically, the error bars represent the asymptotic standard error associated with the fit parameter $d$, corresponding to the oscillation frequency $\omega_2$.

We can conclude that the CG simulations are in much better agreement with the analytical solution than the SC simulations. A contributing factor to this could be the significantly higher spurious velocities present in the SC case, as was shown in section \ref{STCGvsSC}. In the simulations performed here we measure oscillations on a scale of less than the spacing of two adjacent grid-points - see Fig. \ref{fig:diameterInTime} - through interpolation and subsequently fitting the simulation data. At this length-scale even relatively minor perturbations - e.g. due to spurious velocities - can therefore significantly affect the results. The physics of an oscillating droplet is captured properly with an acceptable error using the CG model. The error is shown to be significantly reduced by increasing the resolution.

\subsection{Ligament contraction}\label{ligcont}

\begin{figure}[h]
\centering
\includegraphics[width=0.75\textwidth]{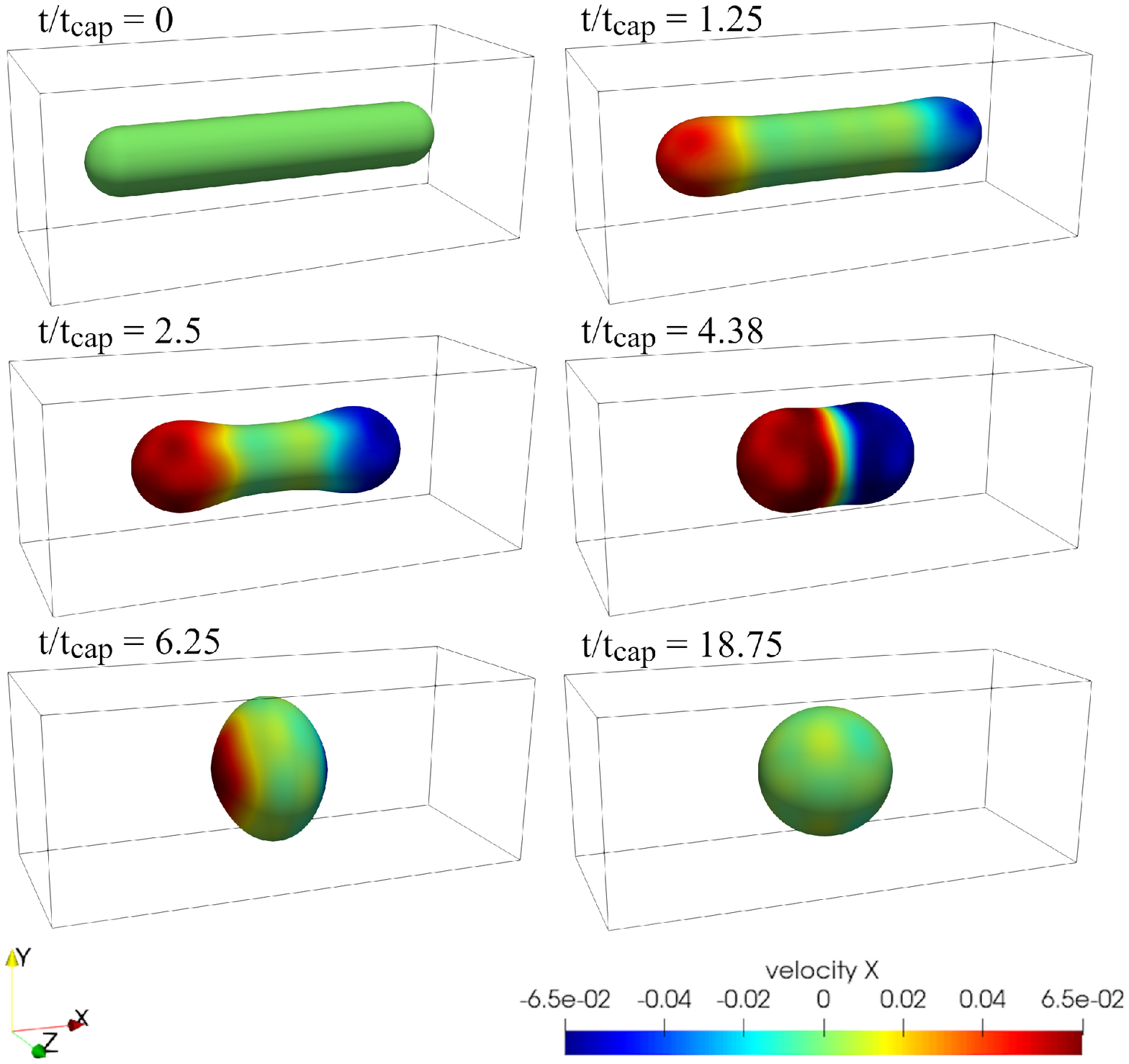}
\caption{\label{fig:contractionVisualization} (CG) Visualization of a typical viscous ligament contraction simulation. Frames shown are at time $t/t_{cap} = 0, 1.25, 2.5, 4.38, 6.25, 18.75$ with $t_{cap} = \sqrt{\rho_R R_0^3 / \sigma} = 160$. The ligament has an initial radius $R_0 = 10$ and initial length $L = 100$. The isosurface is drawn at $\rho^N = 0$ and colored by the x-component of the velocity, with the x-axis being along the length of the domain.}
\end{figure}

To further investigate the physical accuracy of both models in a dynamic system we consider the contraction of a fluid ligament due to surface tension. A capsule shaped ligament (i.e. a cylinder with hemispherical ends) is initialized in a domain of size $  L_x \times L_y \times L_z =  128 \times 64 \times 64 $ with periodic boundary conditions applied on all sides of the domain. The position of the endpoint of the ligament is tracked by determining the interface position at each timestep. An example visualization showing the contraction of the ligament in time is presented in Fig. \ref{fig:contractionVisualization} where a sequence of snapshots shows the contraction into a spherical droplet. An analytical approximation for the endpoint position of the ligament as a function of time is given in \cite{SRIVASTAVA_2013} where it is derived that the endpoints travel with velocity $ u = \sqrt{\sigma / (\rho_R R_0)} $. The timescale considered is the capillary time $t_{cap} = \sqrt{\rho_R R_0^3 / \sigma} $ with initial ligament radius $R_0$.  It should be noted that the analytical approximation reported in \cite{SRIVASTAVA_2013} is based on a force balance where any influence of the surrounding fluid is neglected, i.e. the ligament is considered to be suspended in an inviscid fluid. To closely mimic this situation we set a low viscosity in our simulation for the ambient fluid, $\tau_B = 0.55$. This contributes to minimize momentum loss due to interaction with the fluid surrounding the ligament.

\begin{figure}[h]
\centering
\includegraphics[width=0.75\textwidth]{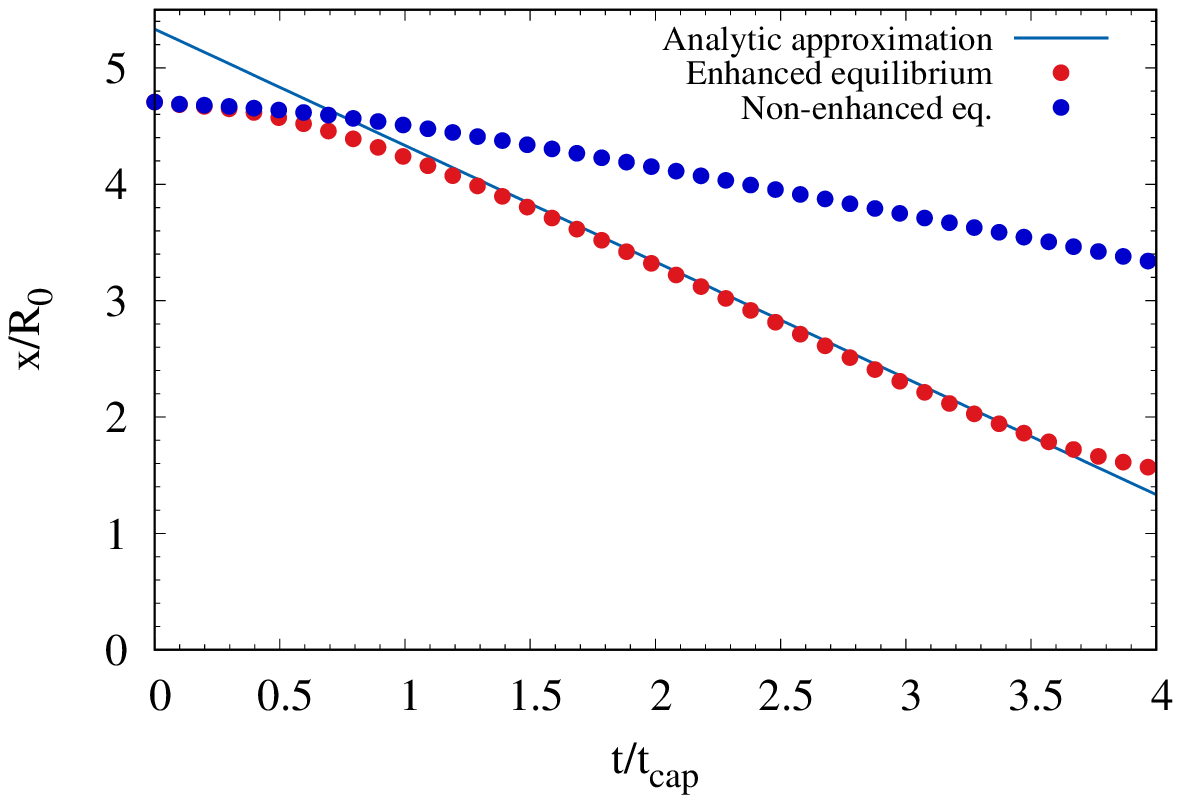}
\caption{\label{fig:ligamentContractionR12} (CG) A ligament is initialized with an initial radius $R_0 = 12$, $\sigma = 0.05$, $\rho_R = 2$, $\rho_B = 0.1$, $\tau_R = 1$ and $\tau_B = 0.55$ (for CG only).  Densities and viscosities are such that $Oh = \mu / \sqrt{\rho_R \sigma \ L} = 0.1$ and the initial ligament length $L = 112$. The ligament will contract due to surface tension and eventually become a spherical droplet. During the contraction phase one of the two endpoints of the ligament is tracked and its position $x/R_0$ is compared to the position predicted by the analytical approximation at time $t/t_{cap}$, with $t_{cap} = \sqrt{\rho_R R_0^3 / \sigma} = 262.9$. After an initial startup phase the rate of contraction is found to be in excellent agreement with the analytical approximation when enhanced equilibrium terms are used. Without enhanced equilibrium terms the rate of contraction deviates significantly from the predicted values.}
\end{figure}

First, we compare two CG simulations, one with enhanced equilibria and one without. The initialization and parameters used are identical with $R_0 = 12$, $\sigma = 0.05$, $\rho_R = 2$, $\rho_B = 0.1$, $\tau_R = 1$, $\tau_B = 0.55$ and the initial ligament length $L = 112$ (reported in Table \ref{table:contrParams} as set 1). These parameters are chosen such that $Oh = 0.1$, using the definition $Oh = \mu / \sqrt{\rho_R \sigma \ L}$, which is a typical value associated with the jetting of microdroplets \cite{dimJet}. This set of parameters results in a capillary time $t_{cap} = \sqrt{\rho_R R_0^3 / \sigma} = 262.9$. The results of both simulations are shown in Fig. \ref{fig:ligamentContractionR12}. It is clear that the standard implementation of CG (without enhanced equilibrium terms) deviates significantly from the analytical solution. In this case the simulated rate of contraction is significantly lower than predicted. However, the case where enhanced equilibrium terms are included shows an excellent match with respect to the analytical solution.

\begin{table}[ht]
\centering
\caption{\label{table:contrParams} Input parameters for ligament contraction simulations: red and blue fluid density $\rho_R$ and $\rho_B$, relaxation times $\tau_R$ and $\tau_B$, surface tension $\sigma$, coupling parameter $G$, initial ligament radius $R_0$ and initial ligament length $L$.}
\begin{tabular}[t]{lccccccccc}
\hline
Set & Method & $\rho_R$  & $\rho_B$   & $\tau_R$    & $\tau_B$   & $\sigma$  & $G$ & $R_{0}$ & $L$ \\ 
\hline
1 & CG  & 2 & $ 0.1 $  & 1.0  & 0.55  & 0.05 & - & 12 & 112 \\ 
2 & CG     & 1.99 & $ 0.165 $  & 1.0  & 1.0  & 0.078 & - & 10 & 100 \\ 
3 & SC    & 1.99 & $ 0.165 $  & 1.0  & 1.0  & - & $-5.0$ & 10 & 100  \\ 
4 & CG     & 1.99 & $ 0.165 $  & 1.0  & 0.55  & 0.078 & - & 10 & 100  \\ 
\hline
\end{tabular}
\end{table}

\begin{figure}[h]
\centering
\includegraphics[width=0.75\textwidth]{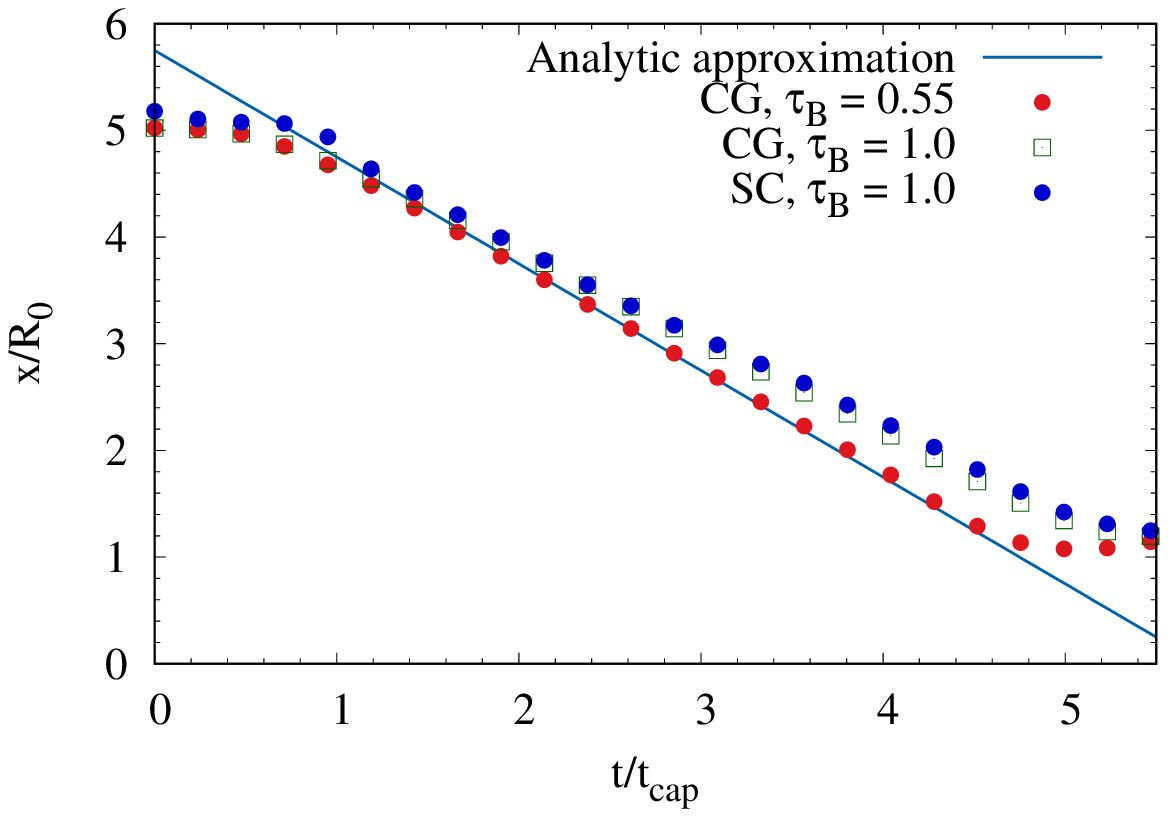}
\caption{\label{fig:ligamentContractionR10SC} (CG vs. SC) A ligament is initialized with an initial radius $R_0 =10$, $\sigma = 0.0778$, $\rho_R = 1.99$, $\rho_B = 0.165$, $\tau_R = 1$ and $\tau_B = 0.55, 1$ and length $L = 100$. The ligament will contract due to surface tension effects and eventually take the shape of a spherical droplet. During the contraction phase one of the endpoints of the ligament is tracked and its normalized position $x/R_0$ (w.r.t. the center of the domain) is compared to the position predicted by the analytical approximation at time $t/t_{cap}$, with $t_{cap} = \sqrt{\rho_R R_0^3 / \sigma} = 160$. After an initial startup phase the rate of contraction is in reasonable agreement with the analytical approximation for both the CG and SC cases with $\tau_B = 1$. Note that the slightly slower contraction compared to the analytical approximation can be explained by the fact that the approximation assumes the ambient fluid to be inviscid. When setting $\tau_B = 0.55$ (giving a less viscous ambient phase) the results match very well with the analytical approximation.}
\end{figure}

We next consider a side-by-side comparison between CG and SC simulations where we pick a set of (stable) parameters from the investigations performed in section \ref{STCGvsSC}. The parameter selection is reported in Table \ref{table:contrParams} set 2 and 3 (corresponding to $Oh = 0.084$ and capillary time $t_{cap} = 160$). In this case the viscosity of the ambient fluid is relatively high, meaning a match with the analytical approximation should not necessarily be expected. The results are shown in Fig. \ref{fig:ligamentContractionR10SC}. A close match is observed between CG and SC simulations with matched parameters. As might be expected, the ligament contraction rate of these two simulations deviates from the analytical approximation due to the relatively high value of $\tau_B$. For the CG model an additional simulation is run where all parameters are kept the same expect for a lower viscosity with $\tau_B = 0.55$, see Table \ref{table:contrParams} set 4. In this case the CG model once again shows excellent agreement with the analytical approximation.

For the ligament contraction case we can conclude that the physics is properly captured using either the CG or SC model for the analyzed range of parameters. It is however necessary to include the enhanced equilibrium distributions to get proper agreement between simulation results and analytical approximation when using the CG model. The SC simulation is in close agreement to the matched CG simulation, which was not the case for the droplet oscillation test considered in section \ref{droposc} where a significant deviation from both the analytical solution and CG simulation was observed. This is likely due to the two cases having a different sensitivity to small disruptions in the system. In the ligament contraction case the relevant movement of the interface is on the order of several grid-spacings, whereas in the droplet oscillation case even movement on a sub-grid length-scale is relevant to determining the oscillation frequency. Small perturbations (e.g. due to spurious currents) will therefore have a smaller effect in the ligament contraction case.

\subsection{Rayleigh-Plateau breakup}

The previous cases have shown the accuracy of both models to capture the dynamics of droplet oscillations and ligament contraction due to surface tension. We will now consider a case that involves interfacial breakup, namely breakup of a ligament into separate droplets due to a Rayleigh-Plateau (RP) instability. An infinite cylinder of fluid is initialized in a domain with  length $  L_x = 128 $ or $ L_x = 256 $ and periodic boundary conditions.

\begin{figure}[h]
\centering
\includegraphics[width=0.75\textwidth]{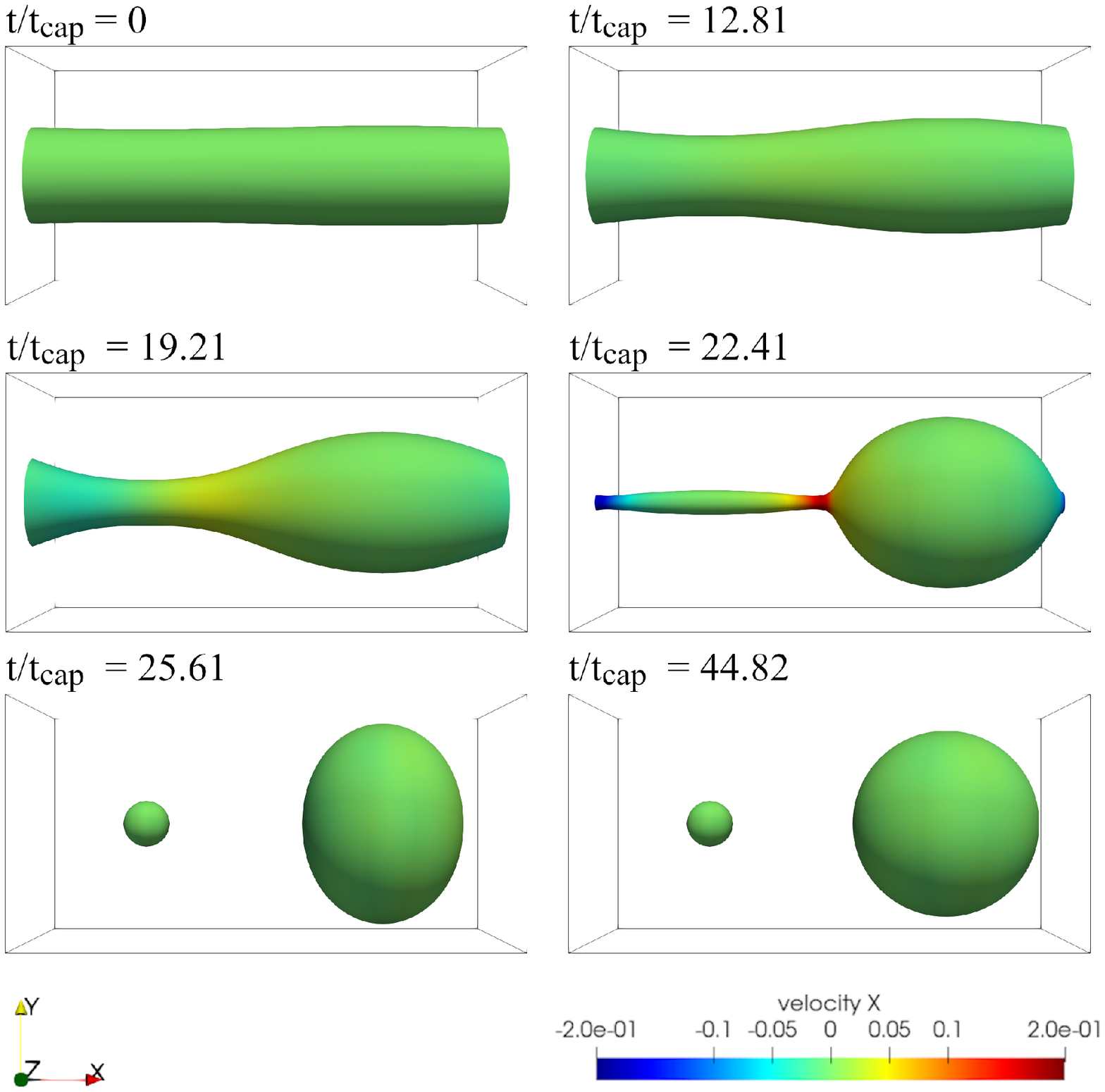}
\caption{\label{fig:RPVisualization} (CG) Visualization of a typical Rayleigh-Plateau breakup sequence. Frames shown are at time $t/t_{cap} = 0, 12.81, 19.21, 22.41, 25.61, 44.82$ with $t_{cap} = \sqrt{\rho_R R_0^3 / \sigma} = 210.2$. Initial radius $R_0 = 12$ with an initial perturbation amplitude $A = 0.5$ and wavelength  $ \lambda = 128 $, equal to the domain length. The isosurface is drawn at $\rho^N = 0$ and colored by the x-component of the velocity, with the x-axis being along the length of the domain.}
\end{figure}

It is expected that whenever the RP instability criterion is satisfied, i.e. $2 \pi R_0 < \lambda$, two droplets will be formed and their size will be dependent on the wavenumber $ \kappa = 2 \pi R_0 / \lambda $, with initial cylinder radius, $R_0$, and perturbation wavelength, $\lambda$. We will refer to the larger of the two droplets as the main droplet and the smaller droplet as the satellite droplet. A visualization showing the breakup process is shown in Fig. \ref{fig:RPVisualization}. In the simulations performed here an initial perturbation is applied with $\lambda = L_x$ and a perturbation amplitude $A = 0.5$.

First a set of simulations using the CG model is run and compared to the slender-jet (SJ) model by plotting our results alongside data taken from \cite{slenderjet}. We consider two sets of simulations, one with $L_x = \lambda = 128$ and one with $L_x = \lambda = 256$ where $R_0$ is selected such that a similar range is spanned for $\kappa$.  Furthermore the ligament density is set to $ \rho_R = 10 $, the surrounding fluid density $ \rho_B = 1 $, $ \sigma = 0.03 $ and $ \tau_R = \tau_B = 0.55 $. These parameters are also reported in Table \ref{table:RPParams} set 1 and 2. The radii of the droplets are measured across the y-axis in the same manner as for the droplet oscillation case, i.e. linear interpolation is used to identify the interface position between two nodes as accurately as possible.

\begin{table}[ht]
\centering
\caption{\label{table:RPParams} Input parameters for RP breakup simulations: red and blue fluid density $\rho_R$ and $\rho_B$, relaxation times $\tau_R$ and $\tau_B$, surface tension $\sigma$, coupling parameter $G$, perturbation wavelength $\lambda$ and initial perturbation amplitude $A$.}
\begin{tabular}[t]{lccccccccc}
\hline
Set & Method & $\rho_R$  & $\rho_B$   & $\tau_R$    & $\tau_B$   & $\sigma$  & $G$ &  $\lambda$ & $A$\\ 
\hline
1 & CG  & 10 & $ 1.0 $  & 0.55  & 0.55  & 0.03 & - &  128 & 0.5 \\ 
2 & CG  & 10 & $ 1.0 $  & 0.55  & 0.55  & 0.03 & - & 256 & 0.5 \\ 
3 & CG  & 1.99 & $ 0.165 $  & 1.0  & 1.0  & 0.078 & -  & 128 & 0.5 \\ 
4 & SC    & 1.99 & $ 0.165 $  & 1.0  & 1.0  & - & $ -5.0$ & 128  & 0.5 \\ 
\hline
\end{tabular}
\end{table}

\begin{figure}[h]
\centering
\includegraphics[width=0.75\textwidth]{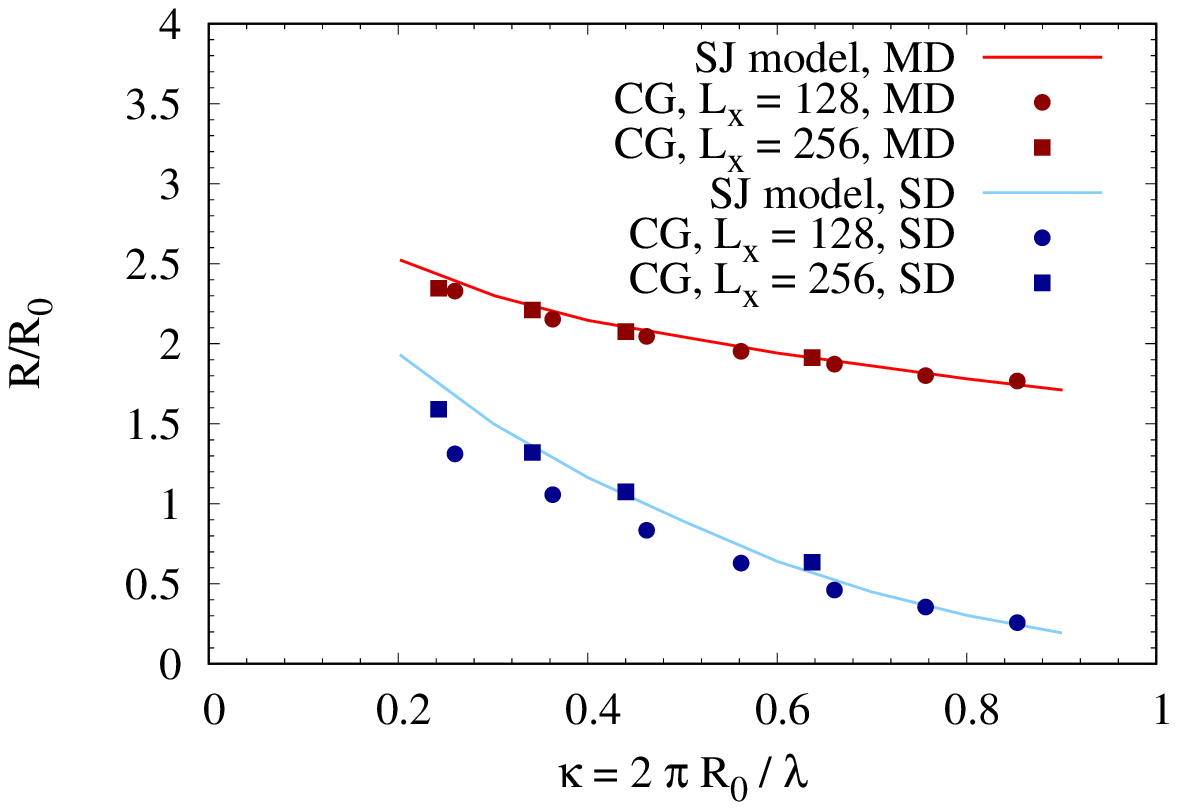}
\caption{\label{fig:radiusVsWavenumberRes} (CG) Rayleigh-Plateau breakup: An infinite cylinder is initialized in a periodic box with domain length $  L_x = 128 $ or $ L_x = 256 $. The ligament density is set to $ \rho_R = 10 $ and the surrounding fluid density as $ \rho_B = 1 $, $ \sigma = 0.03$ and $ \tau_R = \tau_B = 0.55 $. The initial perturbation has a wavelength $ \lambda = 128 $ or $ \lambda = 256 $ and amplitude $ A = 0.5 $. The wavenumber is defined as $ \kappa = 2 \pi R_0 / \lambda $. The initial radius $ R_0 $ is changed to change the wavenumber  $\kappa$ in several runs. Since the RP instability criterion is satisfied, i.e. $2 \pi R_0 < \lambda$ two droplets will be formed and their sizes are dependent on $\kappa$. Simulation results (filled circles and squares) are compared to sizes for the main droplet (MD) and satellite droplet (SD) predicted by the slender-jet (SJ) model \cite{slenderjet}. High resolution simulations show better agreement with the SJ model, especially for low $\kappa$.}
\end{figure}

In Fig. \ref{fig:radiusVsWavenumberRes} the results from simulations with $L_x = \lambda = 128$ (filled circles) are compared to the results with $L_x = \lambda = 256$ (filled squares). In the latter case the initial radius, $ R_0 $, must be doubled to keep $\kappa$ approximately constant between the two simulations. There is an overlap in the wavenumber range for these two setups, which gives information on the effect of increasing the resolution, i.e. increasing $ R_0 $ while keeping $ \kappa $ constant. We conclude that changing the resolution mainly affects the satellite droplet size and in particular for lower wavenumbers. For the main droplets there is little difference in $R$ for the two different resolutions. Both cases are in excellent agreement with the SJ model concerning the main droplets. For the satellite droplets the higher resolution simulations show significantly better agreement with SJ results.

\begin{figure}[h]
\centering
\includegraphics[width=0.75\textwidth]{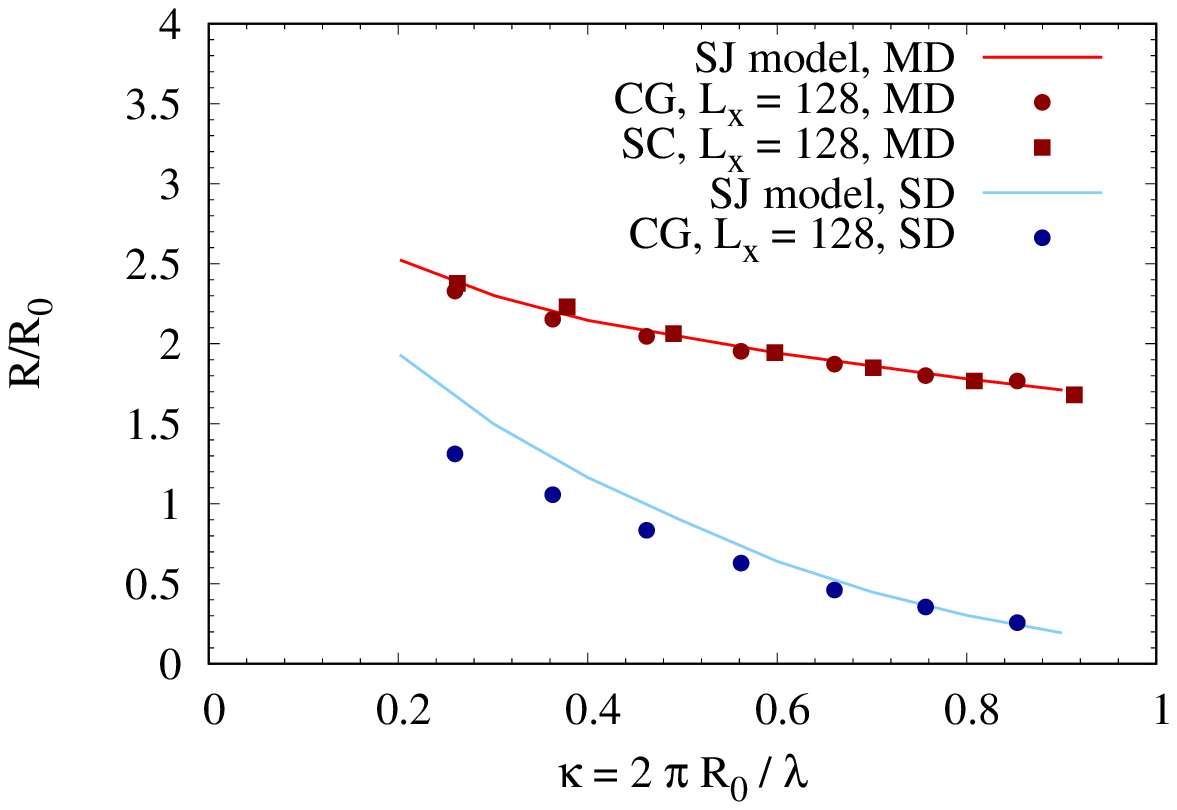}
\caption{\label{fig:radiusVsWavenumberCGvsSC} (CG vs. SC) Rayleigh-Plateau breakup: An infinite cylinder is initialized in a periodic box with domain length $  L_x = 128 $. The ligament density is set to $ \rho_R = 1.99 $ and the surrounding fluid density as $ \rho_B = 0.165 $, $ \sigma = 0.0778$ and $ \tau_R = \tau_B = 1 $. The initial perturbation has a wavelength $ \lambda = 128 $ and amplitude $ A = 0.5 $. The wavenumber is defined as $ \kappa = 2 \pi R_0 / \lambda $. The initial radius $ R_0 $ is changed to change the wavenumber  $\kappa$ in several runs. Since the RP instability criterion is satisfied, i.e. $2 \pi R_0 < \lambda$ two droplets will be formed and their sizes are dependent on $\kappa$. Simulation results (filled circles and squares) are compared to sizes for the main droplet (MD) and satellite droplet (SD) predicted by the slender-jet (SJ) model \cite{slenderjet}. Note that no satellite droplets are present for SC due to rapid numerical evaporation.}
\end{figure}

The next set of simulations is run with the aim of comparing CG and SC side-by-side. We will consider only the lower resolution case with $L_x = \lambda = 128$. The parameters are identical to the side-by-side comparison for ligament contraction, summarized in Table \ref{table:RPParams} set 3 and 4. The results are reported in Fig. \ref{fig:radiusVsWavenumberCGvsSC}. The SC results are in excellent agreement with both CG and SJ models concerning the size of the main droplets. However, all satellite droplets are measured to have a radius $R = 0$, meaning they dissapear rapidly after breakup. This is a consequence of the well known tendency of the SC model to exhibit numerical evaporation or ``mass leakage" \cite{massLeak}. This ``mass leakage" from the droplet to the ambient fluid, where total mass is conserved, is particularly noticeable and occurs more rapidly for relatively small droplets. The CG model does not exhibit this numerical evaporation effect and even small droplets are stable. 

To conclude, the breakup occurs in a physically correct manner for the CG model as the main droplet size as a function of $\kappa$ is as expected when compared to the SJ model. Similarly the satellite droplet sizes are also in good agreement with the SJ model, with the agreement being particularly good for the higher resolution case. The SC model also shows good results for the main droplets, but the smaller satellite droplets are not correctly captured by the model due to rapid numerical evaporation. This shows that for an application like inkjet printing, where very small droplets can be formed under specific circumstances \cite{Fraters2018InkjetP}, the CG model is the preferred choice.

\subsection{Droplet collision}\label{sec:dropcol}

In this section we aim at demonstrating the usability of our proposed method of incorporating repulsion forces at an interface in the CG model - as presented in section \ref{sec:repforce} - and compare the repulsion behavior to a SC simulation with matched parameters. We first consider the case of two droplets colliding head-on. The droplets are initialized with radius $R$ and separated by a distance $\Delta x$ along the x-axis, measured between the centers of both droplets. Each droplet is given an initial velocity $|v_x|$ along the x-axis towards each-other. A repulsion force acts at the interface of the droplets and no other (body)forces are applied. For a given repulsion strength (tunable through the parameter $\Pi$ in our proposed model) the repulsion force acts to prevent droplet coalescence when $|v_x|$ is too low to overcome the repulsion upon droplet impact. At higher $|v_x|$ coalescence can occur when the kinetic energy is sufficient to overcome the repulsion force. The repulsion force for SC is incorporated through a multirange potential as described in section \ref{sec:SC}. We aim to compare the behavior both qualitatively, by visualizing the interface at different points in time $t$, and quantitatively by measuring the radius of the fluid bridge, $R_C$, as a function of time $t$ in the cases where coalescence occurs. For the SC simulations a multi-component model is used (whereas previous cases presented in this work used a multi-phase model). 

\begin{table*}[ht]
\centering
\caption{\label{table:dropcolParams} Input parameters for droplet collision simulations: red and blue fluid density $\rho_R$ and $\rho_B$, relaxation times $\tau_R$ and $\tau_B$, surface tension $\sigma$, repulsion strength parameter $\Pi$, attraction coupling parameters $G^{a}_A$ and $G^{a}_B$, repulsion coupling parameters $G^{r}_A$ and $G^{r}_B$, the cross coupling constant $G_{AB}$ and initial droplet radius $R_0$.}
\begin{tabular}[t]{lccccccccccccc}
\hline
Set & Method & $\rho_R$  & $\rho_B$   & $\tau_R$    & $\tau_B$   & $\sigma$  & $\Pi$ & $G^a_{A}$ & $G^a_{B}$ & $G^r_A$ & $G^r_B$ & $G_{AB}$ & $R_{0}$ \\ 
\hline
1 & SC   & 1.18 & $ 0.18 $  & 1.0  & 1.0  & - & - & $-9.0$ & $-8.0$ & $8.1$ & $7.1$ & 0.405  & 14 \\ 
2 & CG  & 1.36 & $ 1.36 $  & 1.0  & 1.0  & 0.023 & $4.05 \cdot 10^{-3}$ & - & - & - & - & - & 14  \\ 
\hline
\end{tabular}
\end{table*}

For the SC case the parameters are taken from \cite{Benzi2009} and reported here in Table \ref{table:dropcolParams} set 1. Using a Laplace law test for a single droplet with $R = 14$ we find the corresponding surface tension $\sigma = 0.023$. Note that unlike the SC multi-phase case, in the multi-component case the total density at any point in the domain is the sum of both densities. This means that at initialization the total density $\rho = 1.36$ is constant at any point in the domain. Therefore in the CG case we set $\rho_R = 1.36$ and $\rho_B = 1.36$ such that at any point in space the total density $\rho = 1.36$, leading to the parameter set shown in Table \ref{table:dropcolParams} set 2.

For the SC model two cases are simulated where (1) no coalescence is observed at impact velocity $|v_x| = 0.06$ and (2) where coalescence is observed at impact velocity $|v_x| = 0.07$. A series of CG simulations was performed to tune the repulsion strength parameter to $\Pi = 0.00405$, yielding a near identical match in behaviour (in terms of droplet movement and deformation) with the SC case.

\begin{figure*}
\centering

\subfloat[Droplets being repelled]{\includegraphics[width=0.8\textwidth]{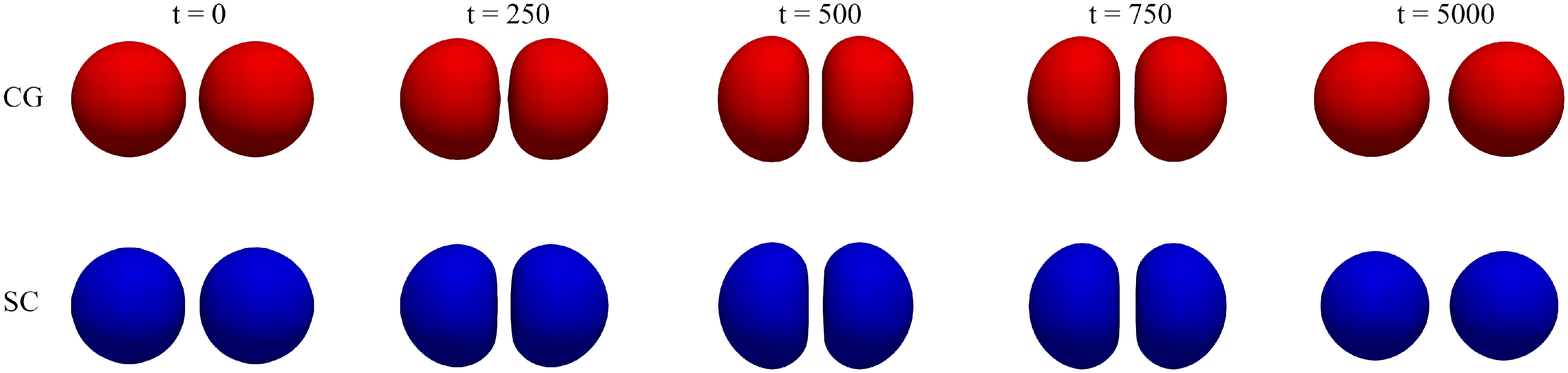}\label{dropRepel}}
  \quad
  \qquad
\subfloat[Droplets coalescing]{\includegraphics[width=0.8\textwidth]{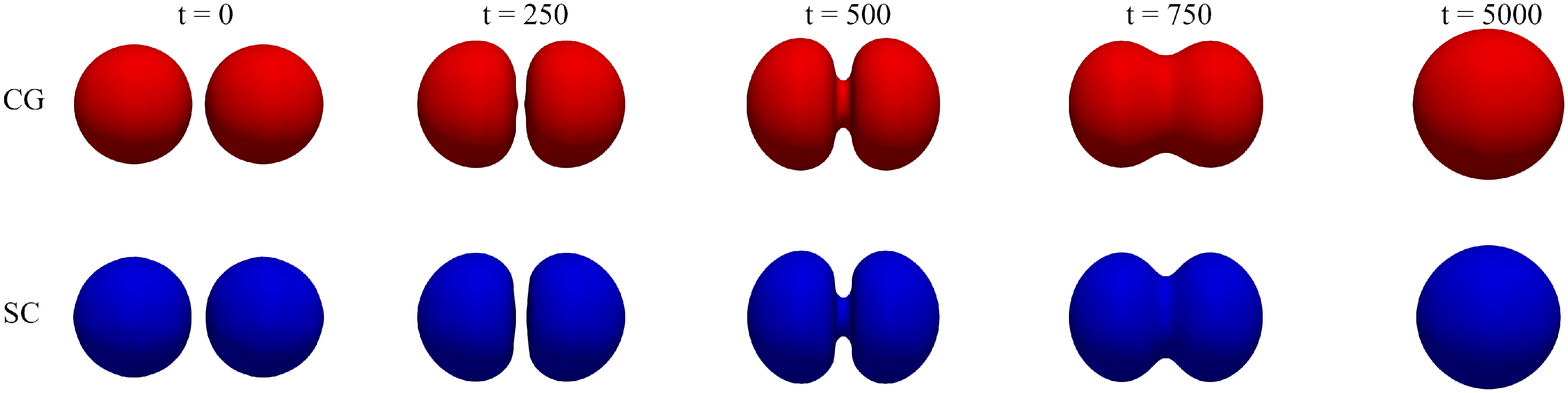}\label{dropCoalesce}}

\caption{(CG vs. SC) Two droplets with radius $R_0 = 14$ are initialized at a distance $\Delta x = 32$ between the centers of the droplets. We set $\rho_R = \rho_B = 1.36$, $\tau_R = \tau_B = 1$ and $\sigma = 0.023$. For the CG simulations the repulsion strength is $\Pi = 4.05 \cdot 10^{-3}$ and for the SC simulations the coupling parameters are set as $G^a_{A} = -9.0$, $G^a_{B} = -8.0$, $G^r_A = 8.1$, $G^r_B = 7.1$ and  $G_{AB} = 0.405$ with $\rho_0 = 0.83$. The CG isosurfaces are colored red, the SC isosurfaces are colored blue. The droplets are initialized with a velocity magnitude $|v_x|$ in the x-direction towards each other. Fig \protect\subref{dropRepel} shows the case with $|v_x| = 0.06$ where the droplets do not have enough velocity to overcome the repulsion force. In Fig. \protect\subref{dropCoalesce} the case with  $|v_x| = 0.07$ is shown, leading to droplet coalescence.}

\end{figure*}

For the qualitative comparison we compare isosurfaces of the interface position. In Fig. \ref{dropRepel} and Fig. \ref{dropCoalesce} we show the repulsion- and coalescence-event respectively. The snapshots taken at time $t = 0, 250, 500, 750, 5000$ show very similar behavior of the droplets for the SC and CG model in the setup we consider here with matched (and tuned) parameters.

\begin{figure}[h]
\centering
\includegraphics[width=0.75\textwidth]{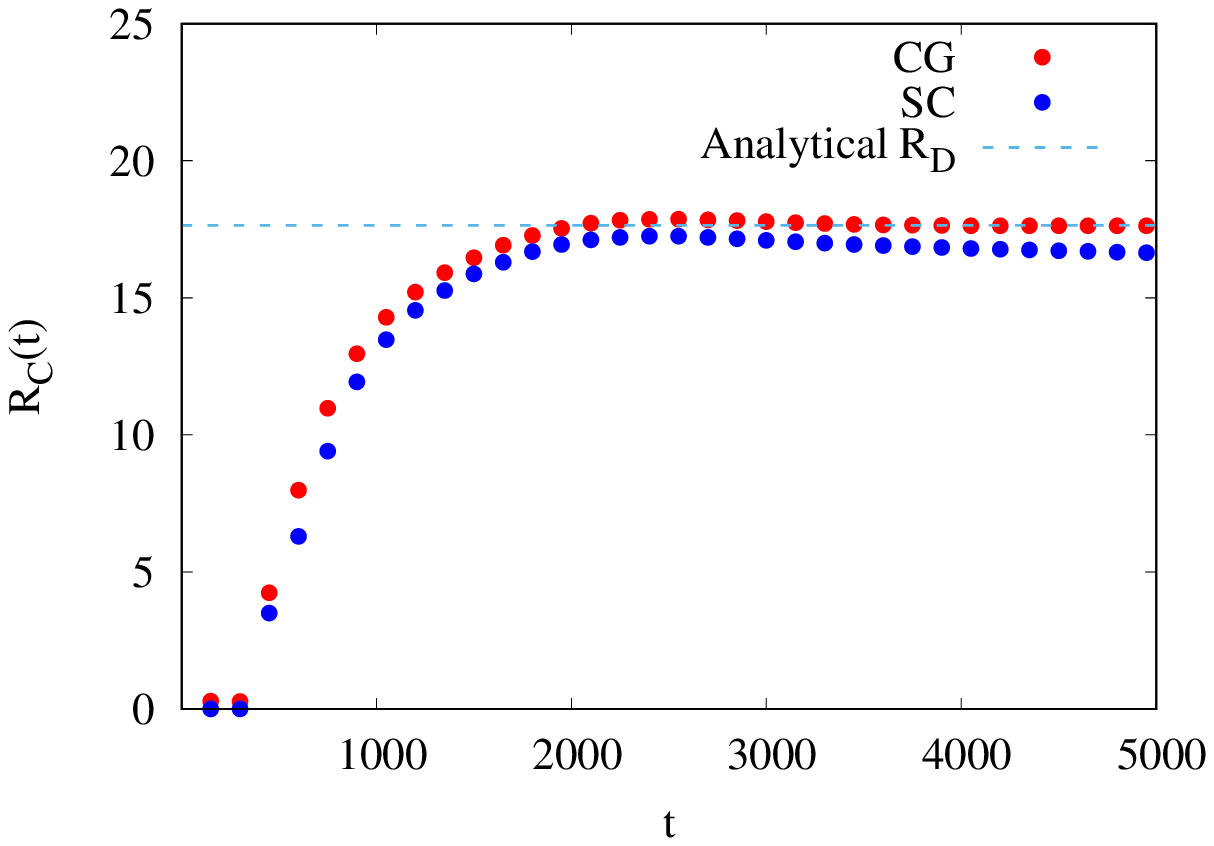}
\caption{\label{fig:diameterInTimeCoalesce} (CG vs. SC) The evolution of the fluid bridge formed during coalescence is quantified by tracking the bridge radius $R_C$ as a function of time. For CG, $R_C(t)$ increases at a slightly faster rate initially, after which $R_C(t)$ becomes very similar for both cases. The equilibrium radius $R_D$ is slightly below the expected value in the SC case, due to mass leakage.}
\end{figure}

A more detailed comparison is possible by measuring the radius of the fluid bridge, $R_C(t)$, that is formed during coalescence at the centerpoint of the domain, see Fig. \ref{fig:diameterInTimeCoalesce}. In the CG model  $R_C(t)$ increases at a slightly faster rate initially, after which the evolution of $R_C(t)$ becomes very similar. A noticeable difference is that $R_C(t)$ is smaller in the SC case over the entire duration of the simulation. This is at least partially due to the mass-leakage effect, which becomes more noticeable as the simulation is run for longer. This is also evidenced by the fact that the eventual $R_C$ at $t = 5000$ is below the analytically predicted equilibrium radius $R_D$ of the coalesced droplet. For the CG model there is an almost exact match with $R_D$, confirming that mass leakage does not occur in this model.

We have shown that for droplet collision at different impact velocities our proposed model is capable of achieving results similar to the more established double-belt potential SC method. The parameter $\Pi$ is able to be freely tuned to match the repulsion caused by the chosen coupling parameters for SC.

\subsection{\label{sec:emul}Emulsion mixing}

To test the repulsion model for a more dynamic and complex system we consider the mixing of two liquids to form an emulsion. Physically the repulsion force can be seen to mimic the effect of a surfactant. During mixing, continuous coalescence- and breakup-events occur under many different impact angles and at many different impact speeds. A domain of size $L_x = L_y = L_z = 96$ is filled with two fluids initialized as planar slabs at a volume fraction $\phi = 0.15$ and stirred by large scale forcing with a stochastic component. Details on the implementation of this force can be found in \cite{massLeak}. We set the same random seed for both the CG and SC simulation, such that the forcing is identical in both simulations. The parameters used in both simulations are reported in Table \ref{table:emulsionParams}.

\begin{table*}[ht]
\centering
\caption{\label{table:emulsionParams} Input parameters for emulsion mixing simulations: red and blue fluid density $\rho_R$ and $\rho_B$, relaxation times $\tau_R$ and $\tau_B$, surface tension $\sigma$, repulsion strength parameter $\Pi$, attraction coupling parameters $G^{a}_A$ and $G^{a}_B$, repulsion coupling parameters $G^{r}_A$ and $G^{r}_B$, the cross coupling constant $G_{AB}$ and forcing amplitude $A_{F}$.}
\begin{tabular}[t]{lccccccccccccc}
\hline
Set & Method & $\rho_R$  & $\rho_B$   & $\tau_R$    & $\tau_B$   & $\sigma$  & $\Pi$ & $G^a_{A}$ & $G^a_{B}$ & $G^r_A$ & $G^r_B$ & $G_{AB}$ & $A_{F}$ \\ 
\hline
1 & SC   & 1.18 & $ 0.18 $  & 1.0  & 1.0  & - & - & $-9.0$ & $-8.0$ & $8.1$ & $7.1$ & 0.405  & $ 5 \cdot 10^{-5}$  \\ 
2 & CG  & 1.36 & $ 1.36 $  & 1.0  & 1.0  & 0.023 & $ 0.012 $ & - & - & - & - & - & $ 5 \cdot 10^{-5}$   \\ 
\hline
\end{tabular}
\end{table*}

\begin{figure*}[h]
\centering
\includegraphics[width=0.8\textwidth]{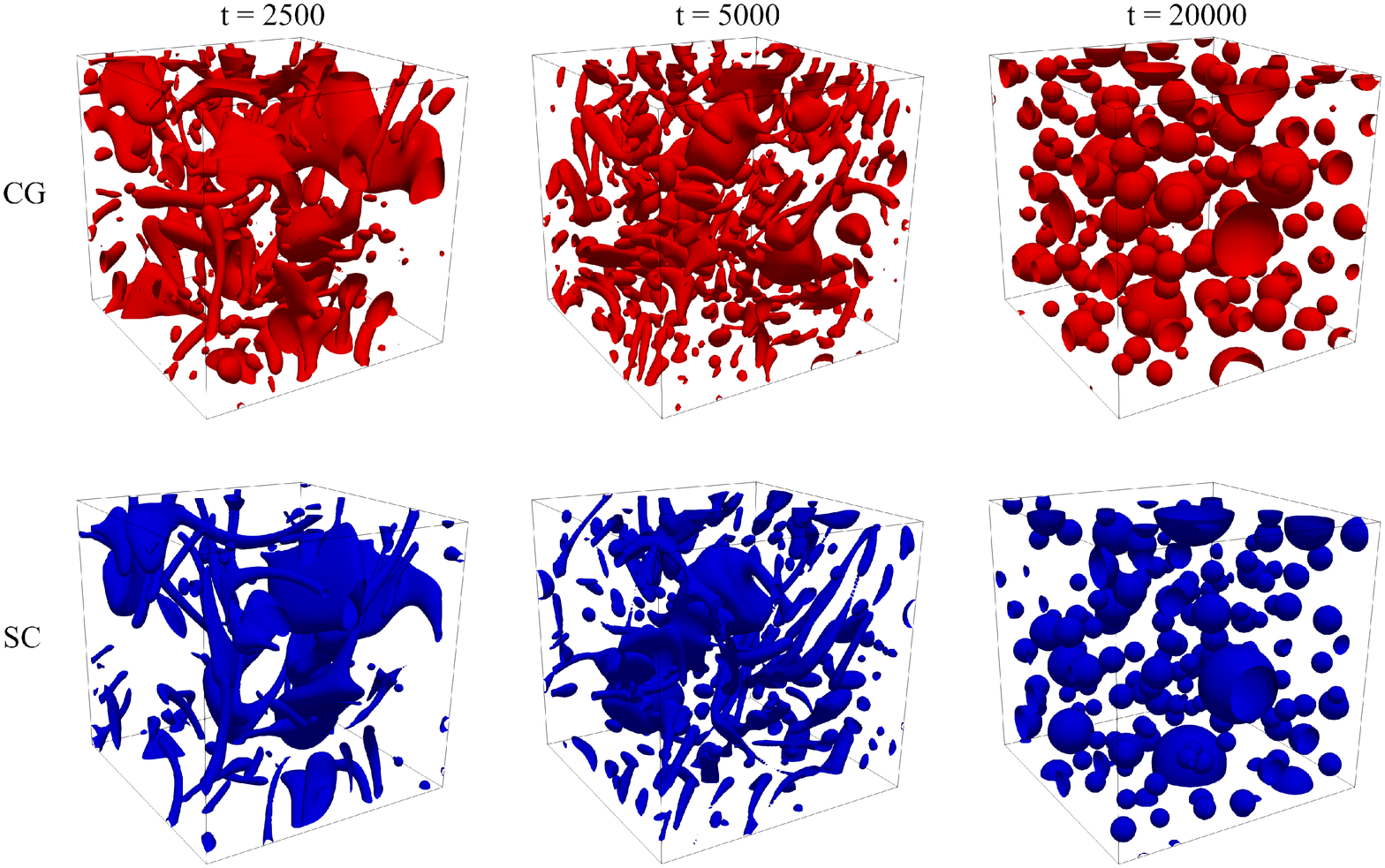}
\caption{\label{fig:emulsionVisualization} (CG vs. SC) Visualization of emulsion mixing with the CG interface colored red and the SC isosurfaces colored blue. Two fluids are initialized as planar slabs at a volume fraction $\phi = 0.15$ at time $t=0$, after which forcing begins immediately, untill timestep $t = 5000$ when forcing is turned off. All other simulation parameters can be found in Table \ref{table:emulsionParams}. Qualitatively both simulations show similar features at the same point in time; at $t = 2500$ mainly sheet and ligament formation has taken place, while at time $t = 5000$ more droplets and thinner ligaments are present in both cases. Finally at time $t = 20000$ the system has relaxed and only spherical droplets are present in the system.}
\end{figure*}

\begin{figure}[h]
\centering
\includegraphics[width=0.75\textwidth]{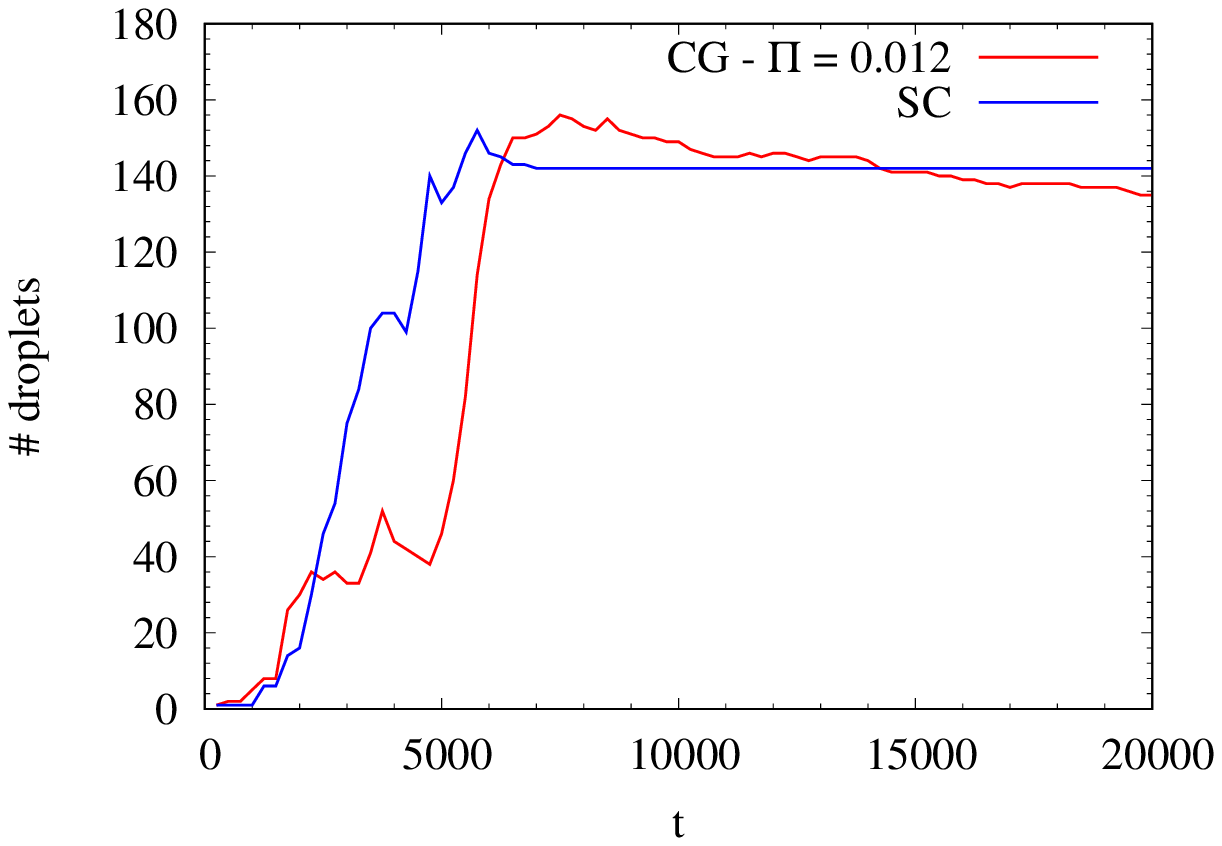}
\caption{\label{fig:dropletsEmulsionCG} (CG vs. SC) Number of droplets as a function of time show for CG and SC. The forcing/stirring is applied for 5000 timesteps after which the forcing is turned off. Some differences in behaviour can be observed, specifically: (1) the delayed formation of droplets which spikes rapidly after $t = 5000$ for CG, whereas in the SC case it is a more gradual increase and (2) the gradual coalescence of (small droplets) after forcing has been turned off up untill the end of the simulation. }
\end{figure}

During and after stirring, the number of droplets is tracked as a function of time. In this case we apply forcing with an amplitude of $A_{F} = 5 \cdot 10^{-5}$ for the first $5000$ timesteps, immediately after which $A_{F} = 0$. The total simulation runtime is $20000$ timesteps, allowing the system to equilibrate and attain a (near) stationary state after the stirring process is complete. All simulation parameters are reported Table \ref{table:emulsionParams}. The repulsion force amplitude $\Pi$ for the CG model was set a posteriori to match the number of droplets formed in the SC multi-component case. 
A qualitative comparison between the two models is shown in Fig. \ref{fig:emulsionVisualization}. Snapshots are taken at times $t = 2500, 5000, 20000$ showing similar (but not identical) structures and features in terms of fluid-ligaments and -sheets being formed. A quantitative comparison is shown in  Fig. \ref{fig:dropletsEmulsionCG} where the number of droplets as a function of time is shown for both methods.  The main notable differences in behaviour between the CG and SC simulations is (1) the delayed formation of droplets which spikes rapidly after $t = 5000$, whereas in the SC case it is more gradual and (2) the gradual coalescence of (small droplets) after forcing has been turned off up untill the end of the simulation. 

We show that the mixing of emulsions can be simulated succesfully with the proposed method of incorporating a repulsion force for CG. Similar results can be attained as in the SC simulation by tuning $\Pi$ although some general differences in behavior are observed.

\section{Conclusions}
The performance of the CG model has been compared to that of the SC model through multiple test-cases. A significantly wider range of parameters is accessible for CG in terms of density-ratio, viscosity-ratio and surface tension values. Numerical stability for a high density ratio of $\mathcal{O}(1000)$ is required for simulating the drop formation process during inkjet printing and is shown to be achievable using the CG model, but not possible using the SC model. The ability to tune surface tension independently from the densities is another major advantage when using the CG model to simulate drop formation during inkjet printing, as a much wider parameter space can be accessed.
In terms of physical accuracy the CG model shows good agreement with analytical solutions for the droplet oscillation and ligament contraction test-cases. The SC model is effective in accurately simulating the case of viscous ligament contraction, but relatively inaccurate in the case of droplet oscillation, which may be caused by the relatively large spurious velocities of the SC model compared to the CG model and the system being sensitive to minor perturbations. It is important to note that inclusion of enhanced equilibrium terms are required to get accurate results in the CG cases as we have confirmed in the ligament contraction case. The RP instability case showed that the CG and SC models give similar results for the measured radii as a function of wavenumber, but only for the larger of the two formed droplets. The smaller satellite droplets quickly vanish in the SC model due to mass leakage towards the ambient phase. This phenomenon does not occur with the CG model, as the scheme enforces immiscibility through the recoloring sub-operator. The absence of mass leakage can be a benefit when simulating inkjet printing. It should be noted that this does mean that (physically realistic) evaporation is not captured in the CG model at all. As long as the amount of evaporation is not significant for the length- and time-scale relevant to a single jetting cycle, the CG model is a suitable choice. 
Finally, we have shown that the repulsion force method introduced in this paper can achieve similar results to the more established SC double-belt potential method by appropriately tuning the tunable parameter $\Pi$. Observed behaviour for the droplet collision test shows a near exact match between CG and SC. The emulsion mixing case also shows a relatively close match in the total number of droplets formed and structures observed during mixing.

The CG model outperforms the SC model in most of the cases considered in this work in terms of stability and accuracy, it does however come at the cost of higher computational demands. The total runtime for a typical droplet oscillation simulation is $486$ s with the SC multiphase model (used for cases A through D). For the comparable simulation using CG the simulation takes $1170$ s to complete, indicating that the SC version is $1170/486 = 2.41$ times faster (the simulations are run on an AMD Ryzen 7 3700X 8-Core processor at 3.60 GHz). This significant difference can be explained by the fact that for the CG model, calculations for both the red and blue fluid need to be made at any one grid-point, effectively doubling the number of calculations required. Additionally the more complex collision operator used in CG requires several additional calculations for evaluating the sub-operators. For the SC multicomponent with multirange potentials model (used for cases E and F) the difference in computation time is less pronounced. The total runtime for a droplet collision simulation is $1780$ s using SC and $2110$ s using CG, indicating that the SC version is $2110/1780 = 1.18$ times faster. This smaller difference is to be expected, since in the SC multicomponent model, similar to with CG, calculations for two fluids need to be made at any one grid-point. Furthermore the addition of the multirange potentials requires additional surrounding grid-nodes (the second ``Brillouin zone") to be evaluated at any one point. 

The computational efficiency of the SC multiphase model may be advantageous for very computationally demanding simulations with a large domain size. This is under the conditions that the parameters lie within the numerically stable range, which is significantly smaller than for CG, and that the reduced accuracy, with respect to CG, is acceptable for the intended application. For the droplet collision and emulsion mixing cases the SC multicomponent with multirange potentials model does not offer the same advantage over CG, as the computation time is only reduced by approximately $20\%$. Furthermore, the CG model with the repulsion model extension presented in this work offers the possibility of exploring a much wider range of parameters, compared to the SC model, allowing its use for the study of dense (stabilized) emulsions.

\section{Acknowledgments}
This work is part of the Netherlands Organization for Scientific Research (NWO) research project High Tech Systems and Materials (HTSM), with project number 13912. The authors thank the NWO and co-financers Canon Production Printing Holding B.V., University of Twente and Eindhoven University of Technology for financial support.

\bibliographystyle{unsrt}
\bibliography{paper}

\end{document}